# High-Throughput Rapid Experimental Alloy Development (HT-READ)

Olivia F. Dippo[a], Kevin R. Kaufmann[b], Kenneth S. Vecchio[a,b]*

**Affiliations:**

[a] Materials Science and Engineering Program, UC San Diego, La Jolla CA 92093, USA.

[b] Department of NanoEngineering, UC San Diego, La Jolla CA 92093, USA

*Corresponding author: kvecchio@eng.ucsd.edu, 9500 Gilman Dr, La Jolla, CA 92093, USA.

**Abstract:** The current bulk materials discovery cycle has several inefficiencies from initial computational predictions through fabrication and analyses. Materials are generally evaluated in a singular fashion, relying largely on human-driven compositional choices and analysis of the volumes of generated data, thus also slowing validation of computational models. To overcome these limitations, we developed a high-throughput rapid experimental alloy development (HT-READ) methodology that comprises an integrated, closed-loop material screening process inspired by broad chemical assays and modern innovations in automation. Our method is a general framework unifying computational identification of ideal candidate materials, fabrication of sample libraries in a configuration amenable to multiple tests and processing routes, and analysis of the candidate materials in a high-throughput fashion. An artificial intelligence agent is used to find connections between compositions and material properties. New experimental data can be leveraged in subsequent iterations or new design objectives. The sample libraries are assigned unique identifiers and stored to make data and samples persistent, thus preventing institutional knowledge loss.

## 1. Introduction

The cycle of materials discovery is highly inefficient, often taking decades for a novel advanced material to be researched, designed, and commercialized [1,2]. Furthermore, much of the information and knowledge generated existed in isolated data silos. This was the motivation for the 2011 Materials Genome Initiative [3], which sparked advances in many high-throughput computational techniques related to materials development [4–7]. However, computational techniques ultimately rely on experimental validation, commonly done using the experimental "one sample at a time" approach; thus, experimental materials research has been critically outpaced by computational advances [8]. As an example of this bottleneck, a case study by Miracle *et al.* found that in the first twelve years of exploration of a new alloy class (multi-principal element alloys), only 122 systems, or $7.2 \times 10^{-7}$ of the possible alloy compositions, were evaluated [9]. At this rate, considering the likelihood of success as equal for every possible composition, the chances of missing the best composition after 100 years of research is 99.9993% [9].

Thus, increasing the rate of materials experimentation is fundamental to improving materials research [8], and requires parallelizing, automating, and miniaturizing key steps in experimental materials research [9], including computation, synthesis, processing, characterization, and data analysis (Fig. 1). High-throughput techniques to accelerate the discovery of functional materials (also sometimes referred to as "combinatorial chemistry") are well-established and reviewed in detail by Potyrailo *et al.* [10]. However, these techniques for the development of functional materials often only apply to 2D materials of interest, and thus fail to capture microstructural and bulk material behavior considerations that are critical in the development of structural materials [9,11–13]. The 3D nature of structural materials introduces additional complications in computation, synthesis, and automation of experimental processes. There has been some effort to develop high-throughput methods for isolated pieces of the materials research cycle for 3D structural materials, for example separate high-throughput synthesis [14,15], characterization [16], and mechanical testing [17–19] techniques; these techniques are well-reviewed by Miracle *et al*. [9] and Boyce & Uchic [20]. However, these techniques, which only optimize one part of the cycle of materials research, inevitably still lead to bottlenecks. To the authors' knowledge, an integrated, closed-loop process of high-throughput alloy development has yet to be demonstrated [2,20].

## 2. Results

*2.1 High-throughput experimental process*

To meet the needs of modern engineering materials design, we have designed an approach and sample geometry that supports the integration of most modeling, processing, and analyses into a streamlined workflow to address the need for high-throughput experimentation for the development of new bulk alloys. To engineer an optimized approach to high-throughput and rapid experimental alloy development, it is essential to develop a methodology that overcomes the most rate limiting steps, and sequentially reduce each remaining rate determining step. Without question, the most time-consuming step in bulk alloy development is the comprehensive microstructural characterization that yields phase fraction, grain size and morphology, phase types, and structural defects. The most comprehensive technique for obtaining and quantifying this data is obtained from the combined techniques of x-ray diffraction (XRD), scanning electron microscope (SEM) based imaging, energy dispersive x-ray spectroscopy (EDS) and electron backscattered diffractions (EBSD). Performing these characterization steps generally requires a dedicated operator, experienced in both physical metallurgy and x-ray/electron microscopy tools, to setup each sample appropriately for the proper data collection, provide decisions regarding phase selection for EBSD as an example, and processing the data. Since techniques like XRD, EDS and EBSD require very specific geometric conditions, establishing a sample geometry that enables automation of these characterization steps is paramount. In this regard, given the geometric constraints of these techniques, there are significant advantages of only employing sample rotations to position from one sample to the next, rather than positioning by X-Y-Z translations. As such, a sample design built around a rotational center of symmetry has significant advantages in terms of automation.

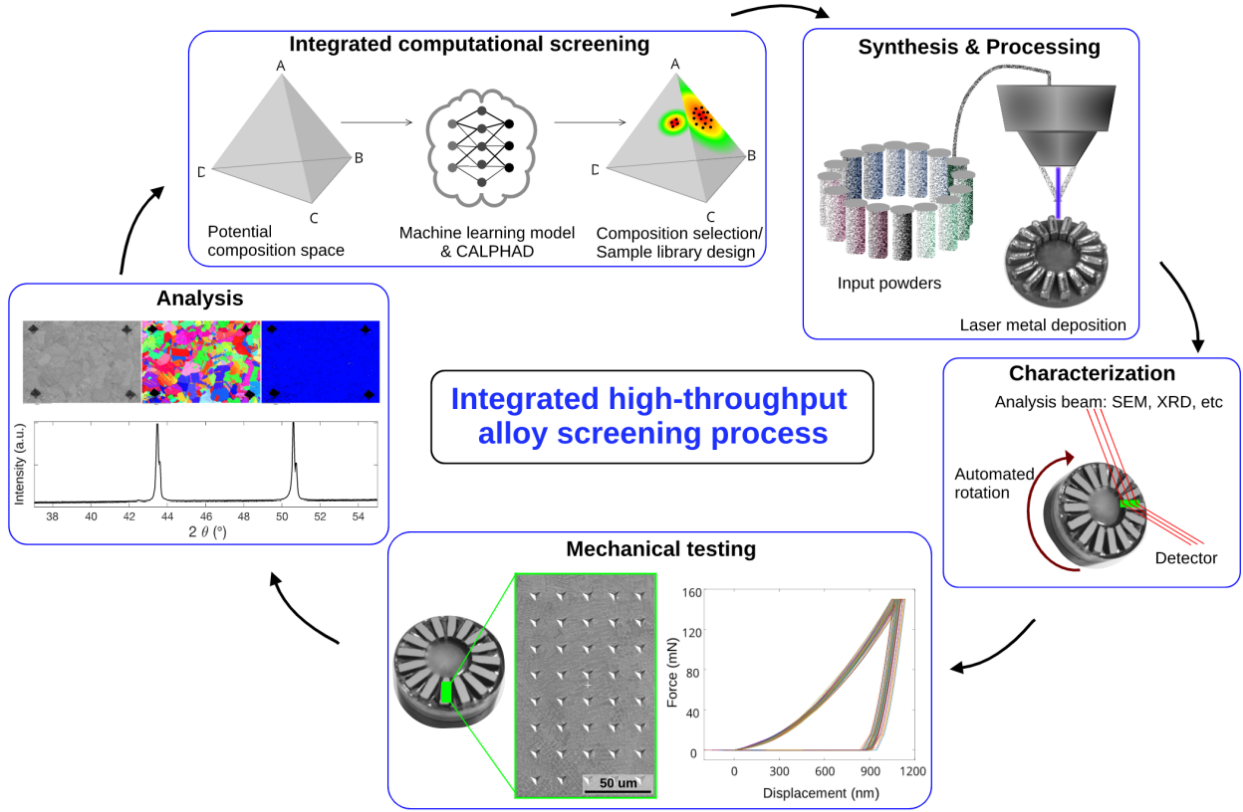

**Figure 1. Illustration of the steps incorporated into the integrated High-Throughput Rapid Experimental Alloy Development (HT-READ) methodology.** Clockwise from top left, computational screening utilizing CALPHAD and machine learning model provide recommendations for sample library compositions. Then samples are synthesized, processed, characterized, tested, and analyzed in an automated, high-throughput fashion. New data is utilized to improve subsequent screening and design.

The miniaturization and automation of experiments described herein begins with the synthesis of a sample "library," as opposed to the traditional one-sample-at-a-time approach. This sample library has been designed to fit in a single metallurgical mount (32mm diameter) and consists of a ring of 16 individual samples (see Fig. 1 and Supplementary Fig. A.1). The library of samples is fabricated using 3D printing, an approach to materials screening suggested in previous works, but not yet implemented for discrete structural alloy screening to our knowledge [21,22]. Previously, 3D printing has been used for building continuous gradient alloys as an approach to high-throughput synthesis [14,23]; however, the inhomogeneity inherent in the gradient approach limits the knowledge gained about microstructure and mechanical properties of specific alloys. The ring configuration with 16 discrete samples allows for individual analysis of compositions while maintaining critical geometries necessary to multiple characterization techniques.

*2.2 Demonstration of the HT-READ process: alloyed Inconel 625*

The ability of the sample array and integrated workflow to foster material innovation, while rapidly and reliably verifying new materials predicted by computation is demonstrated herein using the Inconel 625 (In625) Ni-based superalloy as a model system. Inconel 625 is a nickel-20 chromium based alloy with additions of Nb and Mo, which provide solid solution strengthening [24], and precipitation hardening due to $Ni_3(Nb,Mo)$ precipitates [25], depending on thermal history. In addition to the model alloy system, several commonly applicable computational, processing, and analysis methods are employed to demonstrate the benefits of the defined sample geometry and show what an integrated workflow might look like (Fig. 1). The equipment and processes presented in this work are not meant to be an exhaustive nor concrete list. It is expected that researchers will tailor or build upon the workflow to fit their material system and design goals.

In this demonstration, the composition space was defined starting from Inconel 625's nominal composition (60% Ni, 22.2% Cr, 4% Fe, 10% Mo, 3.8% Nb by weight) and adding up to an additional 13% Nb, 13% Mo, or 19% Cr. Sample 1 was the base Inconel 625 alloy. Samples 2-6 had increasing amounts of Niobium, up to an additional 13 weight percent. Samples 7-11 had increasing amounts of Molybdenum, up to an additional 13 weight percent. Finally, samples 12-16 had increasing amounts of Chromium, up to an additional 19 weight percent. The additive build of these 16 alloy compositions was done using the Directed Energy Deposition (DED) technique in conjunction with the Alloy Development Feeder (ADF), an automated 16-vial powder feeder as illustrated in Figure 1 and shown in Supplementary Movie A.1. Characterization of AM powders used can be found in Supplementary Figure A.2, and for more details on synthesis, refer to Materials and Methods.

Specific sample compositions were selected on the basis of calculated phase diagram (CALPHAD) modeling with the goal of increasing solid solution strengthening, and then gradually forming secondary phases (see Appendix B). Figure 2 exhibits some of the data collected in a high-throughput manner from this demonstration of one sample library, including: crystal structure information gleaned from XRD, composition from EDS, and mechanical properties from instrumented micro- and nano-indentation (Supplementary Figs. A.3 to A.5). Hardness increases with increasing amounts of alloying elements are achieved (Fig. 2C), as well as the gradual transition from a single-phase material to forming multiple phases (Fig. 2D), as expected in this first demonstration of the HT-READ process. A composition list with CALPHAD predicted and experimentally present phases in each composition can be found in Supplementary Table A.1.

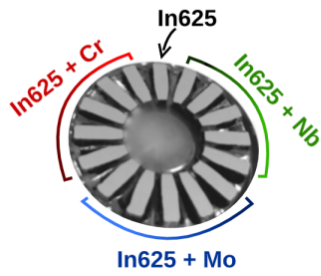
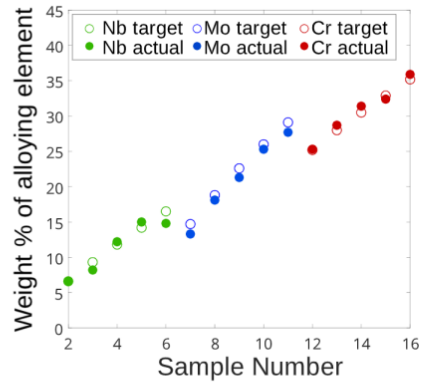
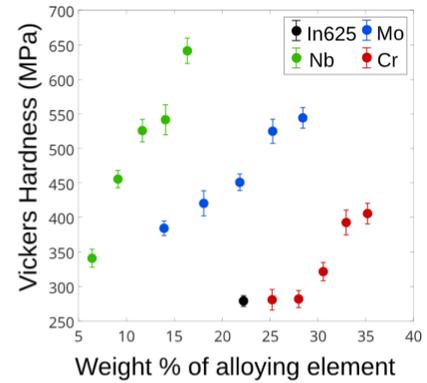
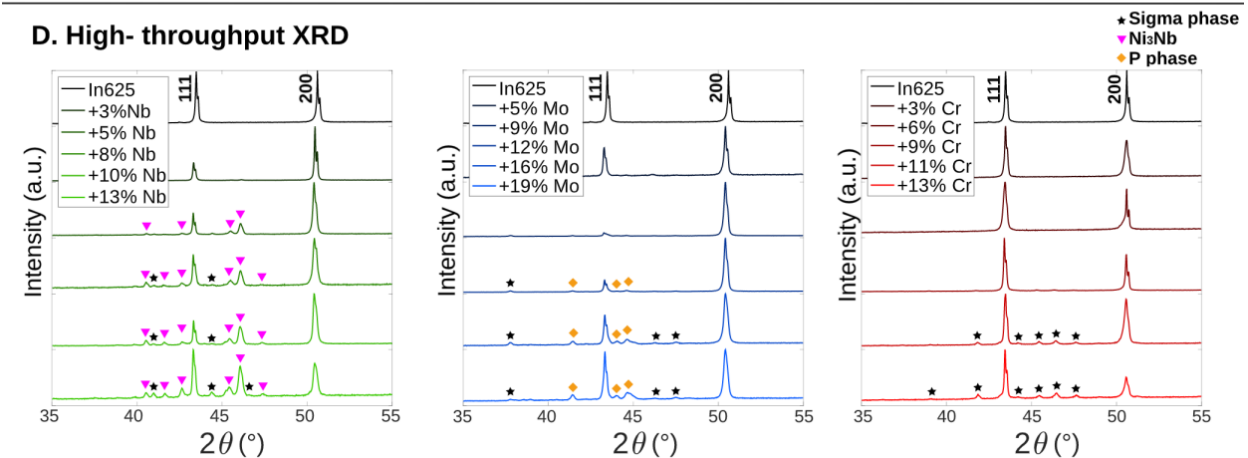

**Figure 2. High-throughput data collection: composition, hardness, and XRD.** (**A**) Sample diagram showing compositions of samples, including In625 base alloy, and three remaining thirds of the sample containing increasing amounts of alloying elements Nb, Mo, and Cr. (**B**) Comparison of target sample compositions (open circles) and actual built sample compositions from EDS (filled circles). (**C**) Measured hardness on In625 and each alloyed variation. (**D**) XRD patterns from the high-throughput process.

Figure 3 exhibits the additional phase information captured from automated SEM-EDS/EBSD, including phase location, size, and morphology for one third of the sample library containing Nb additions (see Supplementary Fig. A.6 for the full sample library). By characterizing the sample with respect to fiducial markers, phase-specific mechanical properties can also be deduced from SEM-EDS/EBSD data (Supplementary Fig. A.3). Due to the ring-shaped sample design, beam geometries in the SEM and the XRD can be maintained as the sample moves on a rotational axis (Supplementary Movie A.2 and A.3; Supplementary Fig. A.7). This allows either the SEM user or a robotic SEM system [26] to set up the first XRD or SEM scan, while the rest of the scans are fully automated. At present, all data collection is completely automated, including the 16 sample XRD, the 16 sample SEM-EDS/EBSD, 16-sample fiducial micro-indentation pattern, and the 16 sample nanoindentation.

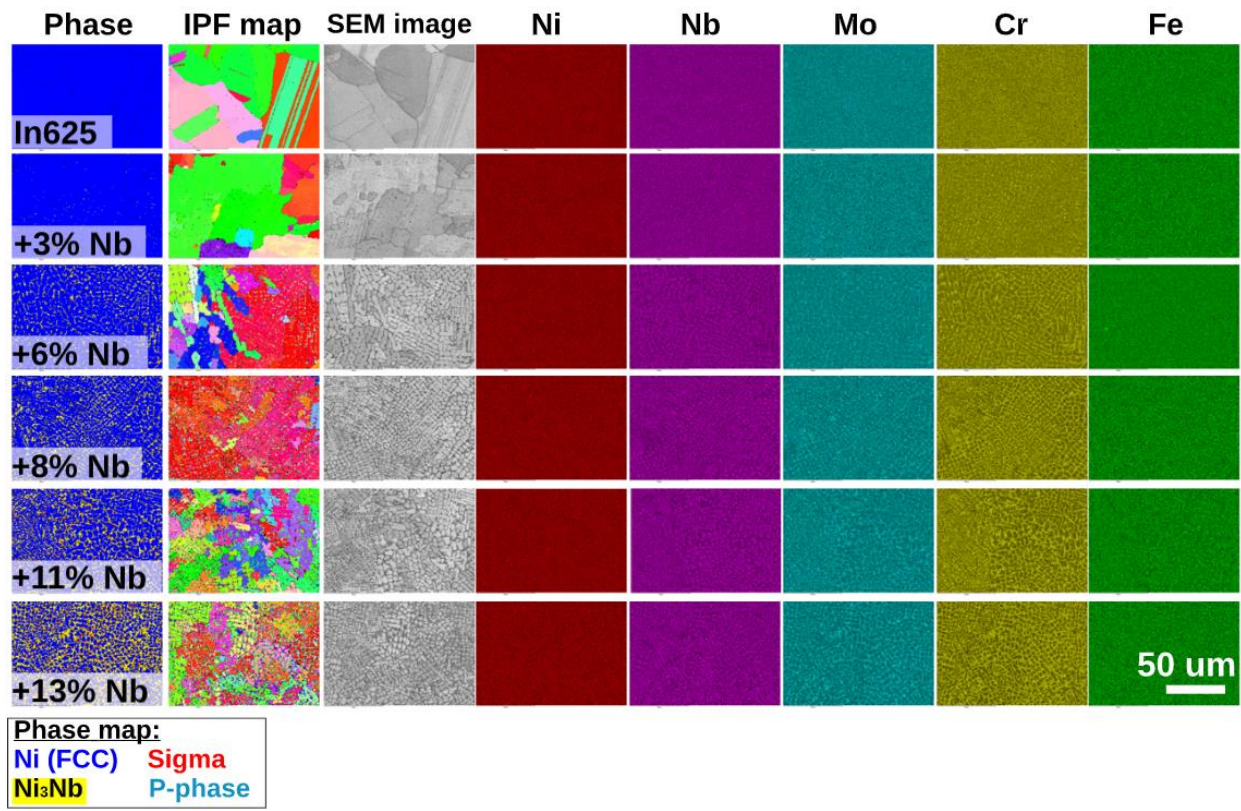

**Figure 3 High-throughput SEM-EDS/EBSD.** SEM characterization for the niobium-alloyed In625, including electron images, phase maps and IPF maps from EBSD, and element distribution maps from EDS. All SEM data is collected in an automated fashion, generating key information about phases present in the microstructure.

*2.3 Machine-learning assisted experiments and hardness prediction*

      As automation has increased within the materials design cycle, data generation rates have exceeded the manual and time-consuming human-driven data analysis process. The development of artificial intelligence (AI) agents offers opportunities to assist researchers from material conception through characterization. Recent works develop and demonstrate these AI agents assisting with composition selection [27–30], property prediction [7,31,32], or chemical and structural analysis [33–36]. Several of these tools could be incorporated into the process, particularly in the characterization steps where data generation rates are rapidly accelerating [36]. For example, AI tools for phase differentiation in EBSD [34,37,38] can be combined with AI enhanced XRD analysis [35] while leveraging chemistry information from EDS to compare phases to existing databases. For some applications, it may be desirable to automate analysis of SEM [33] or optical [39] micrographs collected from each sample in the library. Similar to the processes and equipment in HT-READ, the AI agents utilized are interchangeable depending on the needs of the researchers. While scientists and engineers hold some skepticism for these black box models, they are of considerable value when they efficiently produce results that meet or exceed state of the art systems, the penalty for an incorrect prediction is low, or new understanding of the field is

elucidated [40]. In the case of material design, these AI agents typically use numerical representations of the materials of interest to construct mappings, from input features to the target properties, thus assisting researchers in selecting the next best suitable candidates. This work demonstrates the use of such an AI's assistance by predicting Vickers hardness *in silico*. Hardness is chosen owing to ease of measurement and its correlations with other material properties that are more challenging to measure such as wear resistance and tensile strength [18,41,42].

The hardness values for Inconel 625 alloy and each of the fifteen variations were predicted leveraging Automatminer, an automated machine learning (autoML) algorithm only requiring compositional information as input [43]. Information from CALPHAD modeling was also provided to the autoML process given its observed benefits to other works [27,44]. In practice, this step would be performed in tandem with other computational work (e.g. CALPHAD); however, this work emphasizes exploration of a broad composition space and intended secondary phase formation. Predicting hardness *in-silico* was therefore an auxiliary goal. The Automatminer pipeline leverages evolutionary algorithms to perform many of the necessary optimization steps typically implemented by trained researchers to automatically construct a series of data transformations and machine learning models (Fig. 4A). After optimizing the model and hyperparameters, the fitted model can be deployed to classify new data. A schema of the results from the current work is shown in Figure 4B. The training and test sets have similar data distributions (mean and standard deviation), and contain, respectively, twenty-four and seventeen commercially available Ni-based alloys (Supplementary Table A.2). Nickel content versus hardness for the known data is plotted to show the weak correlation and need for a more advanced model (Supplementary Fig. A.8).

The best score for each of the regression algorithms tested in this pipeline resulted in the selection of a Gradient Boosting Regressor algorithm (Supplementary Fig. A.9 and Supplementary Fig. A.10; Supplementary Table A3). Predictions for the seventeen commercial materials in the test set are plotted to further validate the model (Supplementary Fig. A.11). The computationally determined hardness for Inconel 625 agrees well with existing literature [24] and the experimentally determined value for samples built with a blue (450nm) laser (96W, 120W, and 150W) (Supplementary Fig. A.5). The measured Vickers hardness of the new experimental compositions is compared with the predicted values, and it is determined that three of the five hardest compositions were correctly identified (Supplementary Fig. A.12) despite the small amount of training data and the presence of multiple phases in most of the samples (Fig. 3 and Supplementary Table A.1). Feature importance extracted from the ML model (Supplementary Table A.4) provides insight into how it evaluated the hardness, thus increasing trust in the otherwise 'black-box' model. The combination of the CALPHAD-derived phase evolution diagrams, electron microscopy studies, XRD, ML-predicted hardness, and measured Vickers hardness can thus be applied to subsequent iterations of this closed-loop process.

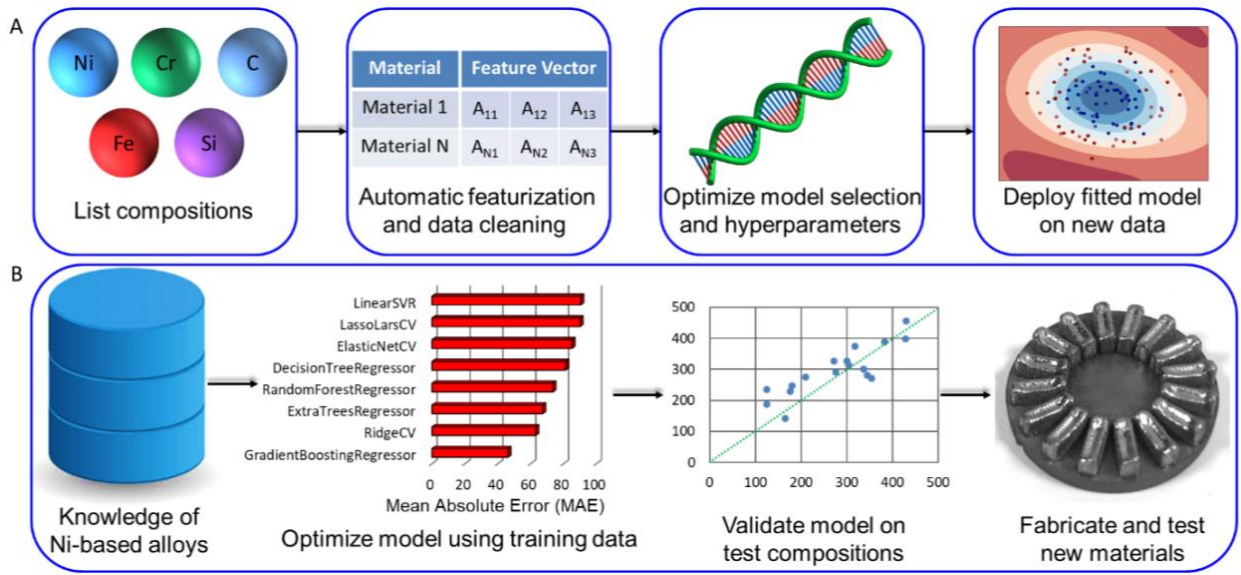

**Fig. 4. Data-driven materials design.** (**A**) Schematic of the machine learning process. The compositional information is transformed into a set of numerical descriptors, a genetic algorithm optimizes the model and its hyperparameters, and the fitted model is tested and deployed on new data. (**B**) Hardness data from 24 Ni-based alloys was utilized to benchmark several models and their hyperparameters. The fitted model is validated with a holdout data set and then applied to select new compositions for experimental testing.

## 3. Discussion

We have presented an integrated high-throughput rapid experimental alloy development (HT-READ) process, leveraging additive manufacturing synthesis, computational screening, and a novel sample library design. Our 16-sample ring library facilitates rapid sample fabrication and automated characterization of phases, microstructures, and properties of structural alloys. Furthermore, the application of 3D printing in this unique "ring" geometry enables materials discovery beyond 2D and functional materials. This pipeline will continue to incorporate AI companion agents to automate and/or accelerate each of the steps. The individual steps in the process can be adapted to different design goals (e.g. conductivity or oxidation resistance) and equipment can be added or removed from the loop to accomplish this. The specifications of the sample library are also readily adaptable to accommodate a variety of equipment. In addition, multiple sample libraries can be fabricated and characterized in parallel. While directed-energy deposition (DED), the additive manufacturing technique used in this work, is currently only amenable to metallic materials, methods exist for additive manufacturing of virtually any class of materials [45]. Most metallic materials are suitable for synthesis by DED, and there are vast compositional spaces for which there is limited knowledge, i.e. high entropy alloys. Using this HT-READ methodology could also lead to the development of new materials that expand the number of ideal materials for commercial additive manufacturing. The HT-READ approach may also lead to more rapid tuning and validation of computational models of materials since many samples are fabricated and tested in a rapid manner, providing validation to a larger number of computations.

There are some limitations and bottlenecks which still exist in the high-throughput experimental process as presented. First, in additive manufacturing synthesis with a revolving powder feeder, vials are loaded with 16 pre-mixed alloy powders, and re-filled with new alloy powders for each new 16-ring sample. However, recent works have achieved *in-situ* powder mixing, by utilizing repositories of individual elemental powders calibrated to flow powder at a rate consistent with desired sample compositions [23,46]. While this in-situ powder mixing approach may have some appeal, expanding the technique beyond 3-4 different elements will face significant hurdles for tuning compositions in higher component spaces. It may be more practical to automate powder weighing and mixing, and subsequent vial filling, in order to retain the composition fidelity needed for complex alloy development. Additionally, researchers currently grind and polish sample surfaces to achieve strict surface quality requirements for nanoindentation and EDS/EBSD characterization using a semi-automated polishing system, although automated sample preparation is commercially available [47]. Uchic *et al.* have designed a fully autonomous system for polishing, transfer, and subsequent SEM characterization [26], and have suggested a clear and imminent path forward for fully autonomous sample handling for alloy development processes [20]. Thus, many of the current limitations of the presented approach can be overcome with existing and impending technological advances. In addition, the flexibility to substitute computational methods and equipment makes the approach accessible and applicable to the scientific community in a variety of materials disciplines.

## 4. Materials and Methods
### 4.1 CALPHAD modeling

Calculated phase diagrams (CALPHAD) is one approach to the modeling of materials [11]. In this demonstration of the integrated approach, CALPHAD diagrams were utilized to predict the resultant phases after printing and heat treatment. CALPHAD calculations were done for each of the individual 16 compositions in the library using the TCNI10 database. For this demonstration, compositions were selected with the goal of gradually increasing the fractions of phases other than the face-centered cubic nickel alloy phase expected in Inconel 625. The CALPHAD results for each of the 16 compositions can be found in Appendix B.

### 4.2 Additive powders

Spherical IN625 alloy powder, particle size 45-90 µm (Oerlikon Metco); -325 mesh Nb powder, (99.8% purity, Alfa Aesar); 30-50 µm spherical Mo powder, (99.9% purity Tekna); and -325 mesh Cr powder (99% purity, Alfa Aesar) were obtained for sample fabrication. 200ppm of an alumina-based flow aid powder (Aeroxide Alu-C, Evonik) was added to each composition to increase flowability through the additive powder delivery system. SEM characterization of all powders used in synthesis, refer to Supplementary Fig. A.2. In625 powder, alloying elements, and flow-aid powder were weighed to achieve desired compositions and blended on a rolling mill before being loaded into designated vials in the DED machine.

*4.3 Sample library design*

The sample library's circular "ring" shape, with an outer diameter of 32mm, was determined by the size and shape of a standard metallographic mount and the respective mounting, grinding, and polishing equipment. The decision to fabricate the samples along the outside ring was motivated by the ability to perform iterative analyses on each sample with minimal alignment (i.e. rotating 22.5-degrees about the Z-axis) after initial geometry setup in most standard lab equipment. X-ray diffraction, energy-dispersive X-ray spectroscopy, and electron backscatter diffraction were used to demonstrate this principle.

*4.4 3D-printing sample libraries*

Samples were individually printed at 22.5-degree angles to one another using a Formalloy Directed Energy Deposition (DED) machine equipped with a 150W 450nm laser produced by Nuburu. The DED machine is also equipped with a Formalloy Alloy Development Feeder (ADF) powder hopper (Movie S1). This powder feeder contains 16 separate vials, designed to rotate via computer control, with one vial feeding into the DED system at a time, allowing for up to 16 distinct alloy compositions within one build cycle. Samples were built on an In625 plate using a laser power of 130-150W and feed rate of 750mm/min. The entire sample library was heat-treated at 1100°C for one hour to solutionize any inhomogeneities resultant from the deposition process.

*4.5 Sample library preparation for analysis*

The 3D printed sample library was cut out in one piece, including the underlying substrate, with a water jet cutter (ProtoMax; OMAX) and mounted in an amorphous resin. The surfaces of the 16 specimens were then simultaneously prepared using standard metallographic procedures: grinding using SiC papers to 4000 grit, polishing using 3μm and 1μm diamond suspensions, and final polishing with a 0.04μm colloidal silica solution.

*4.6 XRD Analysis*

To accommodate the 16-sample ring, an XRD sample holder (Supplementary Fig. A.7) has been designed to facilitate sample rotation and thus automation of the scans in a benchtop Rigaku Miniflex unit. This sample holder ensures the alignment of the flat, polished sample with the diffraction plane of the goniometer, and the beveled glass cover ensures no contamination of signals from neighboring samples in the sample library. This provides a low-cost solution to retrofit XRD systems that do not have sample rotation capabilities.

A Panalytical X'Pert Pro was utilized to demonstrate the capacity for automatic rotation and data collection for each sample in the 3D printed library. All measurements were collected using a 1D detector and Cu Kα radiation (wavelength λ = 1.54059 Å). XRD measurements were taken from 20 to 100 degrees (2-theta), with a step size of 0.02, and scan speed of 12 seconds per step. The sample is then rotated about its omega axis (i.e. "wobble") in 2-degree steps for a total of 5 scans per sample. The 5 scans are then averaged together to create one XRD plot. This "wobble" function reduces texturing effects which can occur due to large columnar grains often present in additive builds.

*4.7 SEM analysis*

Several analyses were performed in a Thermo Scientific (formerly FEI) Apreo FEG scanning electron microscope (SEM), equipped with an Oxford MAX energy-dispersive X-ray spectrometer (EDS) and an Oxford Instruments Symmetry electron backscattered diffraction (EBSD) camera. The EBSD detector used is a CMOS detector, a fast, high-resolution camera capable of diffraction pattern rates up to 3000Hz. EBSD patterns collected were used to generate maps of orientation and phases, allowing determination of key microstructural information including phase fraction, location, and relative strain information in addition to the phase chemistry information provided by EDS. Elemental maps using EDS were collected simultaneously with EBSD, reducing the analysis time per sample to approximately 1 minute.

EBSD requires a large sample tilt (typically 70 degrees, but a 60 degree tilt was used in this work) relative to the SEM pole piece; this geometrical requirement elucidates aspects of the sample ring design. In order to automate this process, the geometry from the pole piece to the sample to the EBSD detector must be maintained. Using a rectangular array of samples at a seventy-degree tilt would require repositioning in x,y, and z, and consequent beam adjustments, i.e. refocusing at each sample. Instead, the 16 sample ring design allows an automated rotation of 22.5 degrees with no motion in x, y, or z directions and no need to refocus. This allows for continuous, automated analysis of all 16 specimens in less than an hour. The combination of these analyses provides information about phase chemistry, phase morphology, phase location, and other details critical to microstructural assessment.

*4.8 Indentation*

Microhardness indentation was used to create fiducial markers approximately 50mm in size, marking a rectangle on each of the 16 samples (Supplementary Fig. A.3). This process is automated with a Duramin-40 automated hardness indenter (Struers). These microhardness indents are used to locate areas of interest during SEM data collection and nanoindentation. Within the microhardness indents, nanoindentation (G200 Indenter, Keysight) with 150mN force was used to create a grid of 40 indents (Supplementary Fig. A.4). The indents were spaced 20mm apart (d), with contact depths between 1-2 mm, maintaining a normalized spacing (d/h value) of greater than or equal to 10 [48]. The fiduciary microhardness indents were used to calibrate the Vickers Hardness obtained by automated nanoindentation (Supplementary Fig.

A.5). Due to indentation size effects, nanoindentation hardness values are often higher than microhardness indentation values. Instrumented nanoindentation with a Berkovich indenter is used experimentally for automation, however the available datasets for machine learning purposes are in Vickers Hardness. Berkovich to Vickers hardness is calculated by multiplying the nanoindentation hardness value by a shape factor of 92.5. Microhardness Vickers hardness values (HV500) from fiducial markings were obtained, and a best-fit line(solid) is used to calibrate nanoindentation hardness values to microhardness Vickers values, subtracting nanoindentation hardness values by 58 HV.

*4.9 Machine learning property prediction*

The Vickers Hardness for the starting Inconel 625 alloy and each of the fifteen variations were predicted using Automatminer, an automated machine learning (autoML) algorithm only requiring compositional information [43]. Information from CALPHAD modeling was also provided to the autoML process given its observed benefits to other works [27,44]. The thermodynamic database utilized was TCNI10. Refer to the GitHub repo for further details of the thermodynamic information. The Automatminer pipeline performs many of the necessary optimization steps typically performed by trained researchers (i.e. feature generation, feature reduction, model selection, and hyperparameter tuning) while leveraging the genetic programming within the TPOT software package [49] to automatically construct a series of data transformations and machine learning models (e.g. random forests, support vector machines, linear models, etc.) with the goal of maximizing performance. The performance of nine regression algorithms and varying hyperparameter combinations are explored via this approach. The mean absolute error (MAE) of each resultant model is recorded, ultimately retaining the best performing model for each type of algorithm (Supplementary Fig. A.10). The Gradient Boosting Regressor algorithm with hyperparameters shown in Supplementary Table A.3 resulted in the best fitting model (lowest MAE) and was therefore selected as the first-generation ML model for predicting Vickers Hardness in Ni-based superalloys. Gradient boosting algorithms are an ensemble of weak sub-models that minimize the loss function by iteratively choosing a function that points opposite the gradient. That is, instead of fitting a new estimator on the data to predict the target variable, the new estimator is trained to predict the residuals of the prior estimator (Supplementary Fig. A.9).

The time necessary for the described autoML process will vary based on the computer hardware, number of data points (training data), and the quantity/quality of the features that represent each composition. In this work, we found this process to complete in approximately 30 minutes employing 1 thread on an Intel Core i9-7920X CPU running at 2.90GHz. The autoML process is thus accessible to the average user and can be accelerated by running jobs in parallel, up to the number of threads on the CPU.

*4.10 Identifiers and database*

Each sample library is given a unique identification number, which is engraved onto the sample ring (but could otherwise be applied by stamping, etching, engraving, etc.). Within each sample library, the individual samples are numbered from one to sixteen. Records of the composition, processing parameters, analyses performed, etc. can be linked to original samples, thus promoting persistence of the data, knowledge, and samples. The choice of database and the information stored within it are ultimately up to the end user of the presented framework.

**Acknowledgements:** We thank Dr. Tyler Harrington for his assistance in the initial conception and design of the high throughput sample array and analysis process. We would also like to thank Grant Schrader for laboratory assistance and Wenyou W. Jiang for aggregating the material property data.

**Funding:** Funding for the development of the customized Formalloy blown-powder DED chamber was obtained through an ONR-DURIP program contract number N00014-16-1-2983. Supported by the U.S. Department of Defense (DoD) [through the National Defense Science and Engineering Graduate Fellowship (NDSEG) Program] and the ARCS Foundation, San Diego Chapter (K.R.K.); and the Oerlikon Group (O.F.D., K.S.V.).

**Author contributions:** O.F.D assisted in developing the idea, performed the bulk of the experimental and analysis work, and led the preparation of the initial draft of the manuscript and figures; K.R.K. assisted in developing the idea, assisted with design and implementation of the 16-sample ring, developed and integrated the machine learning technologies, and assisted with preparing the initial draft of the manuscript and figures; K.S.V. conceived and led the development of the vision for achieving the HT-READ methodology, guided the focus of the project, and reviewed and revised the manuscript; and all authors participated in analyzing and interpreting the final data and contributed to the discussions and revisions of the manuscript.

**Competing interests:** The authors declare no competing interests.

**Data and materials availability:** All data and models generated and/or analyzed during the current study are available from the corresponding author upon reasonable request. All code and models generated, developed, and/or utilized are available at GitHub address https://github.com/krkaufma/htread.

# Appendix A: Supplementary Materials for

## High Throughput Rapid Experimental Alloy Development (HT-READ)

Olivia F. Dippo, Kevin R. Kaufmann, Kenneth S. Vecchio

**This PDF file includes:**

    Supplementary Figures A.1 to A.16
    Supplementary Tables A.1 to A.4
    Captions and stills for Supplementary Movies A.1 to A.4

**Other Supplementary Materials for this manuscript include the following:**

    Supplementary Movies A.1 to A.4
    Appendix B: CALPHAD Data

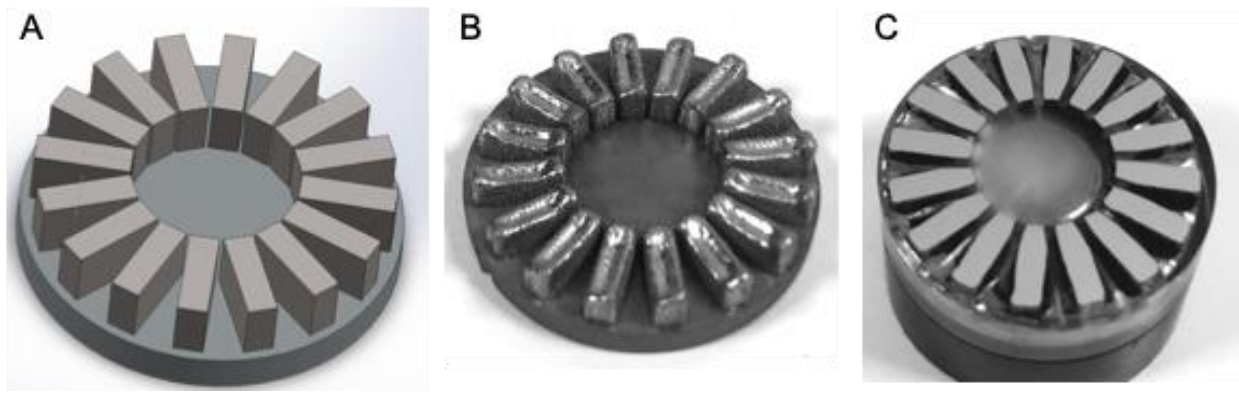

**Supplementary Figure A.1**
**Conception through fabrication of the sample library design.** (**A**) 16-sample library design drawing. (**B**) Additively manufactured 16-sample library. (**C**) Mounted and polished 16-sample library. The diameter of the ring is 1.25 inches, a standard size for automatic metallographic preparation and characterization.

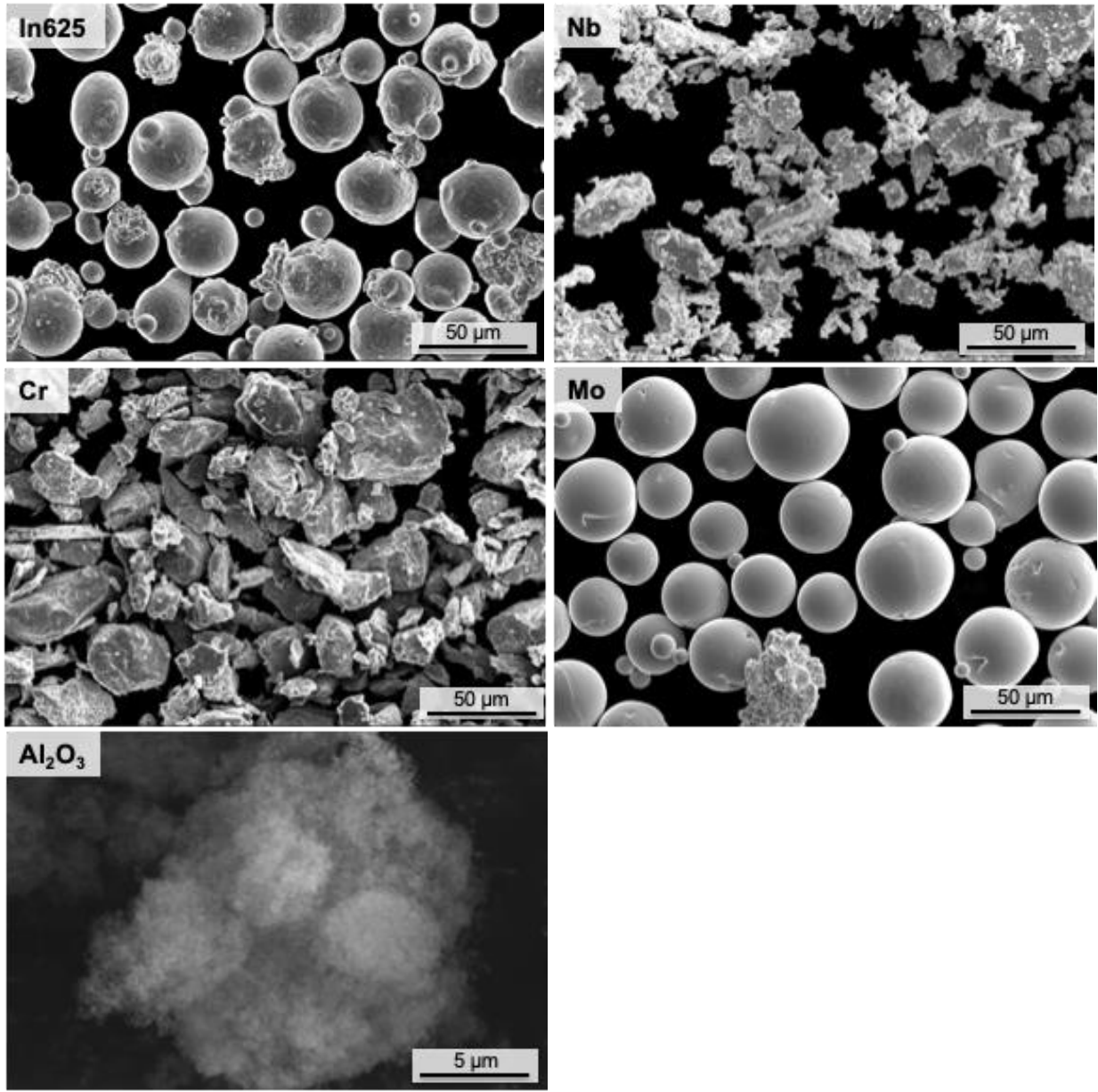

**Supplementary Figure A.2**

**SEM characterization of AM powders.** In625: Inconel 625 alloy (Oerlikon Metco) Nb: -325 mesh niobium powder (Alfa Aesar). Cr: -325 mesh chromium powder (Alfa Aesar). Mo: 30-50µm spherical molybdenum powder (Tekna). Al₂O₃: alumina flow aid powder, AEROXIDE Alu-C (Evonik). Note Al₂O₃ image is at a different scale.

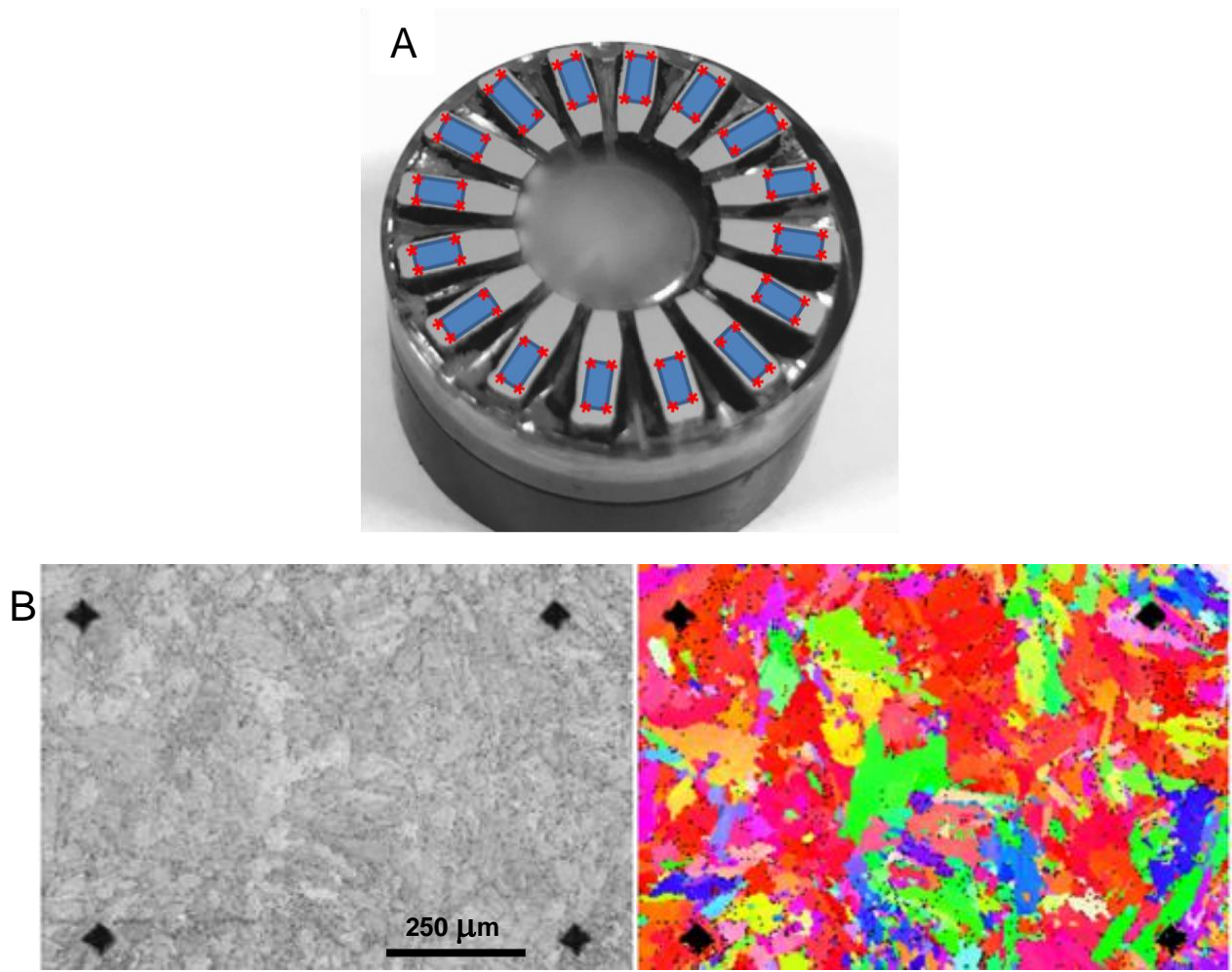

**Supplementary Figure A.3**

**Fiducial markers.** (A) Schematic of fiducial markers made on each of the 16 samples, not to scale. The scale of the area marked for analysis depends on the scale of microstructural features analyzed. The fiducial markers are made using an automated indentation pattern (see methods). (B) SEM images within the fiducial markers: electron image (left) and EBSD IPF map (right). By syncing the automated fiducial marking patterns, automated SEM, and automated nanoindentation, it is possible to extract phase-specific properties from a particular microstructure in a high-throughput manner.

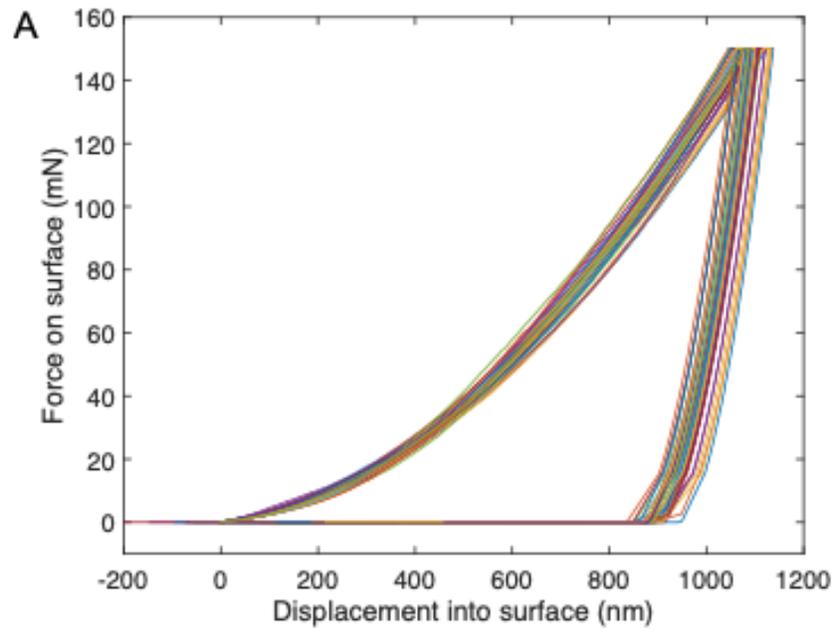

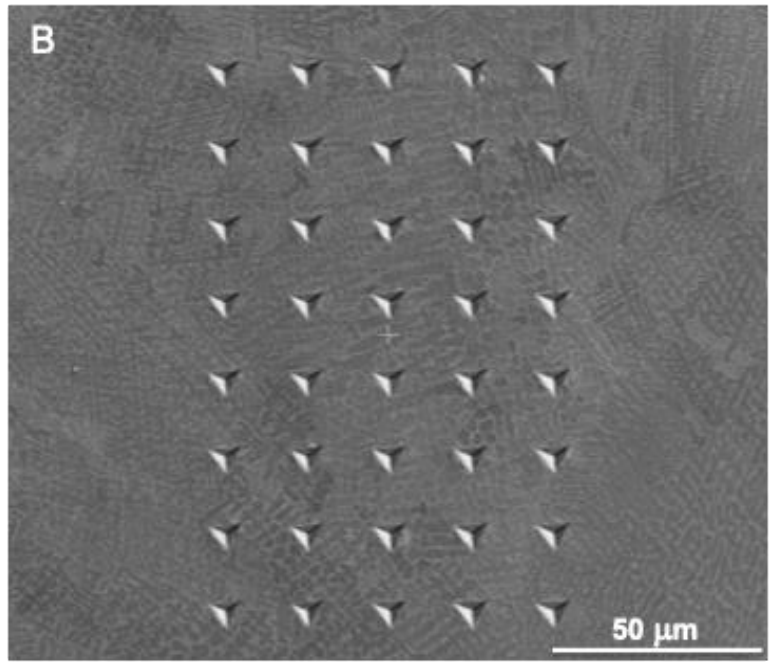

**Supplementary Figure A.4**
**Example hardness characterization for sample 5: In625 + 11% Nb.** (A) Force-displacement curves for 40 indents done at 150mN. (B) SEM image of the grid of 8x5 indents.

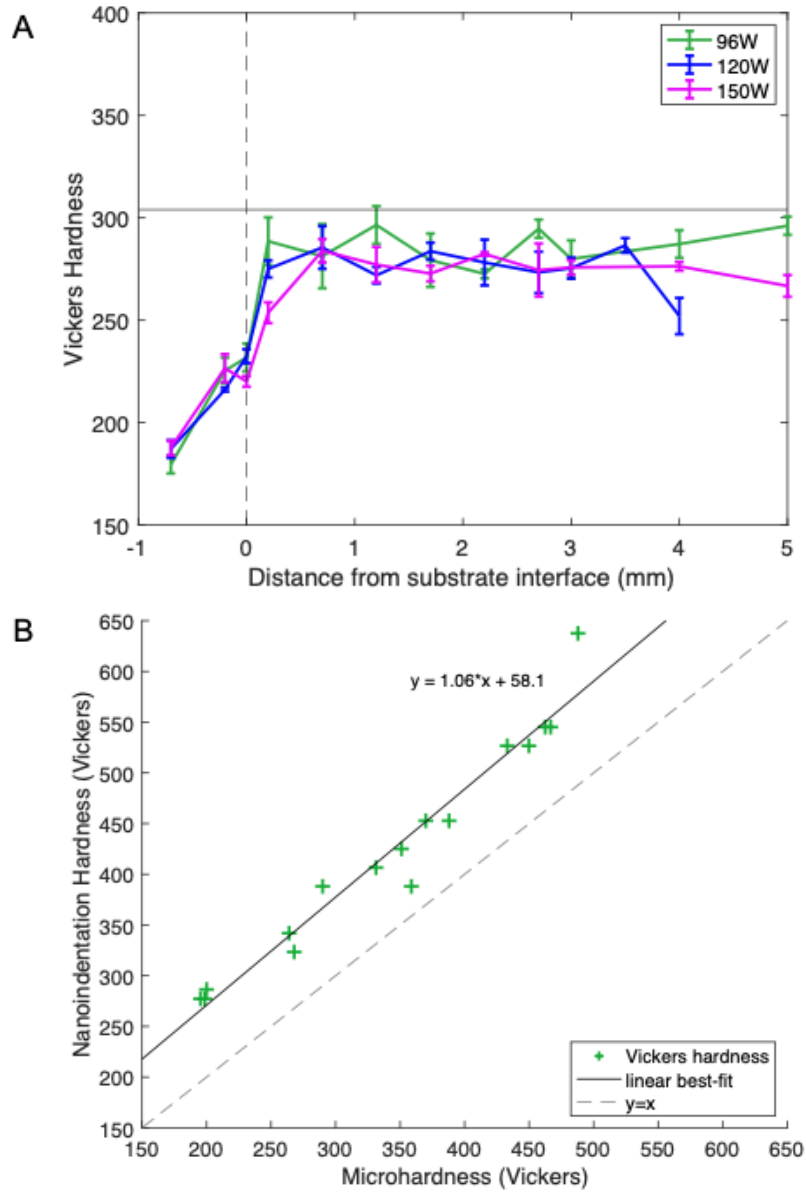

**Supplementary Figure A.5**

**Hardness characterization.** Hardness values are given in HV units. (A) Vickers hardness (HV300) as a function of distance from substrate. Hardness above 200μm from substrate is relatively constant with respect to blue laser power and distance. (B) Microhardness calibration of nanoindentation hardness. Dashed line represents y=x. Linear best-fit line is used for calibration (see Materials and Methods).

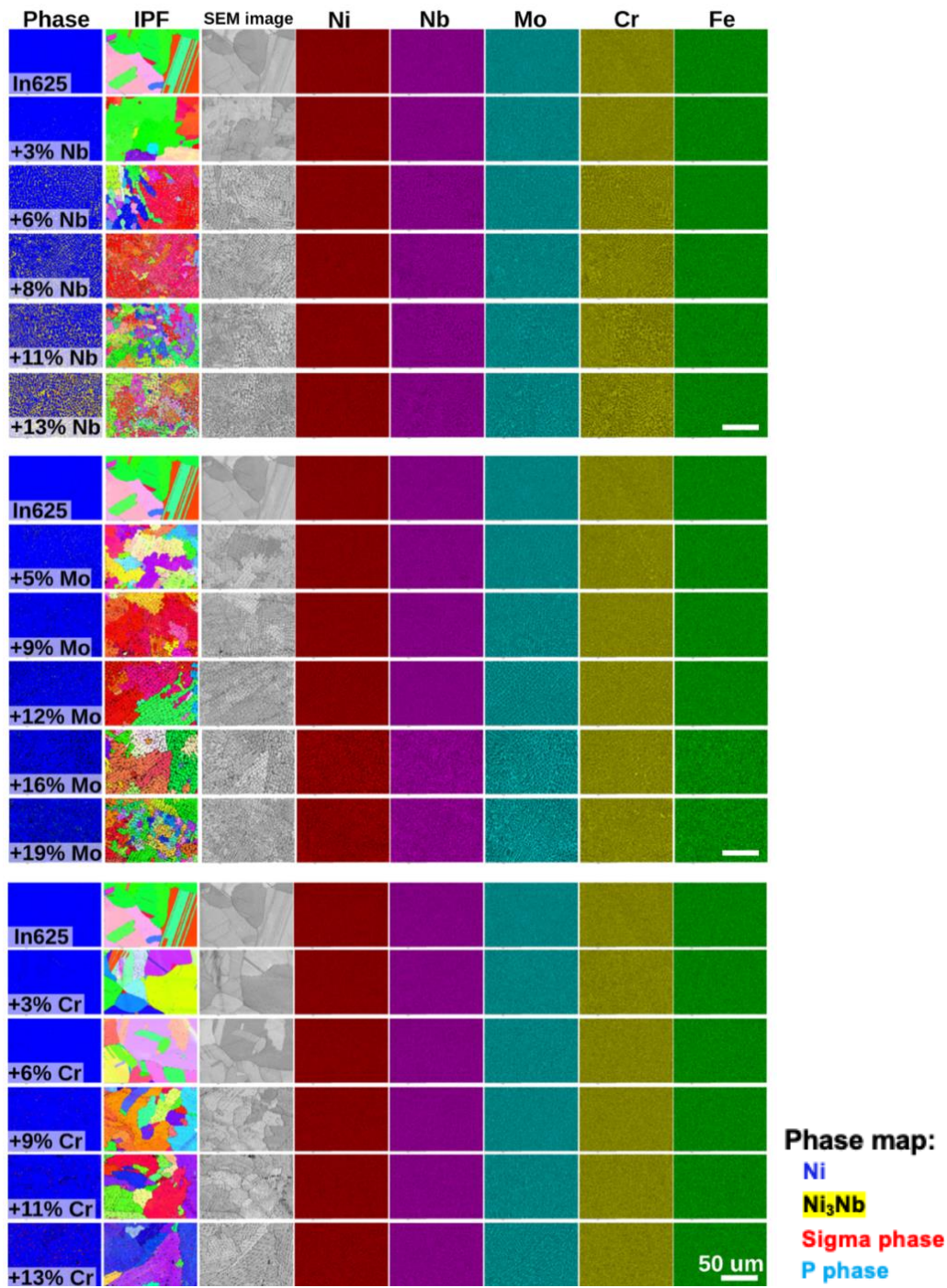

**Supplementary Figure A.6**

**High-throughput SEM data.** EBSD and EDS data from automated, high-throughput SEM data collection. Each block of six samples starts with the base In625 sample at the top and visually details the incremental addition of one element.

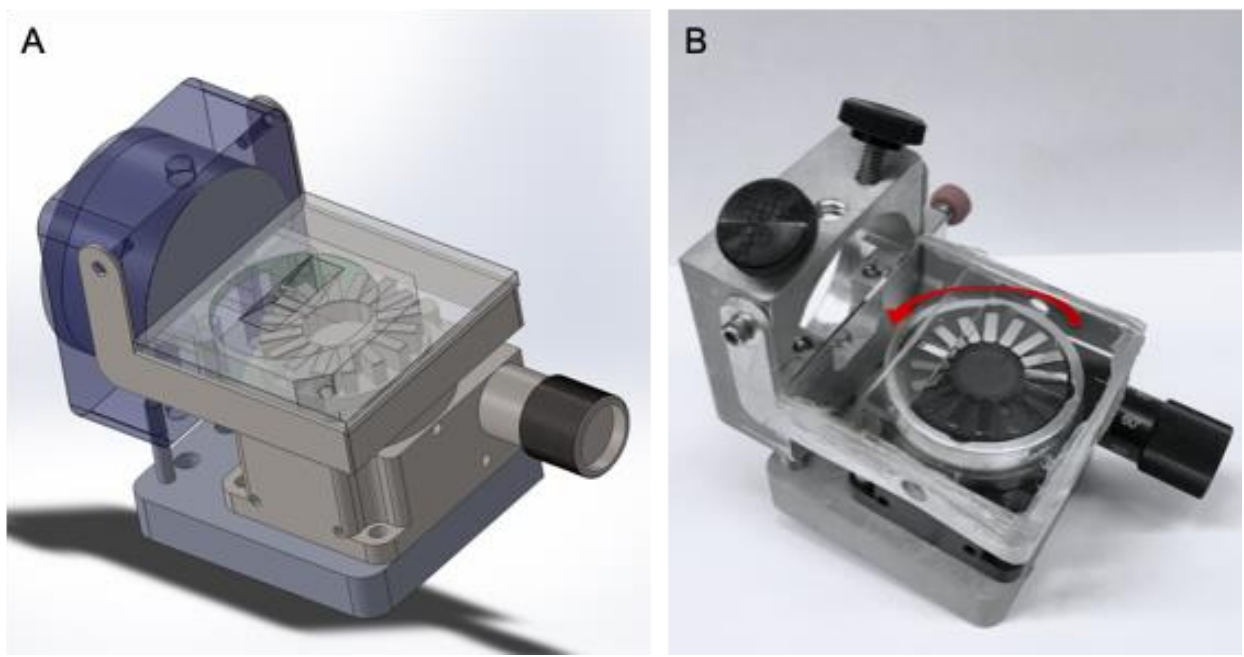

## Supplementary Figure A.7
**Rotational XRD sample holder.** (A) Drawing of the device. (B) Fabricated and tested sample holder for retrofitting benchtop XRDs without sample rotation capabilities. The glass cover ensures that there is no data contamination from neighboring samples.

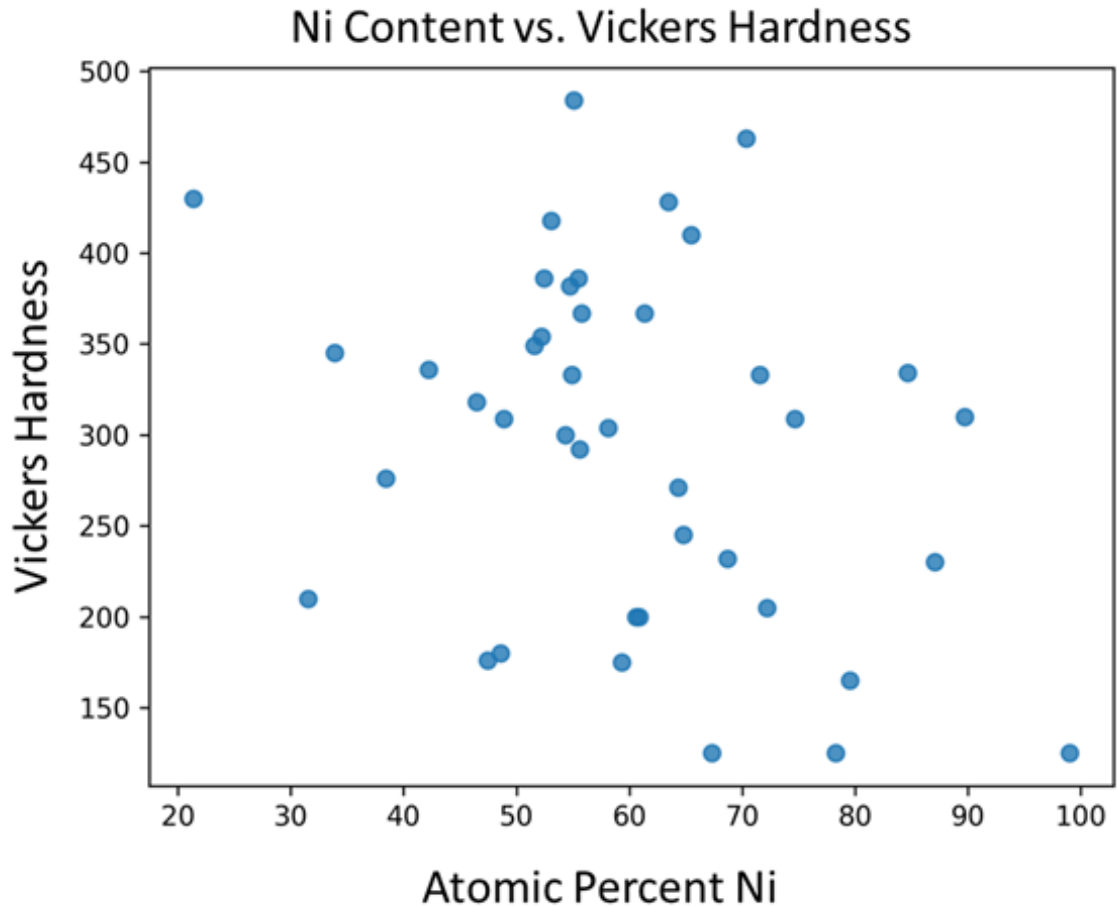

**Supplementary Figure A.8**
**Nickel content and Vickers hardness.** The nickel content (at%) is plotted against the experimental Vickers hardness for materials in the training and test set. The correlation is observed to be weak and therefore not an ideal standalone predictor of hardness.

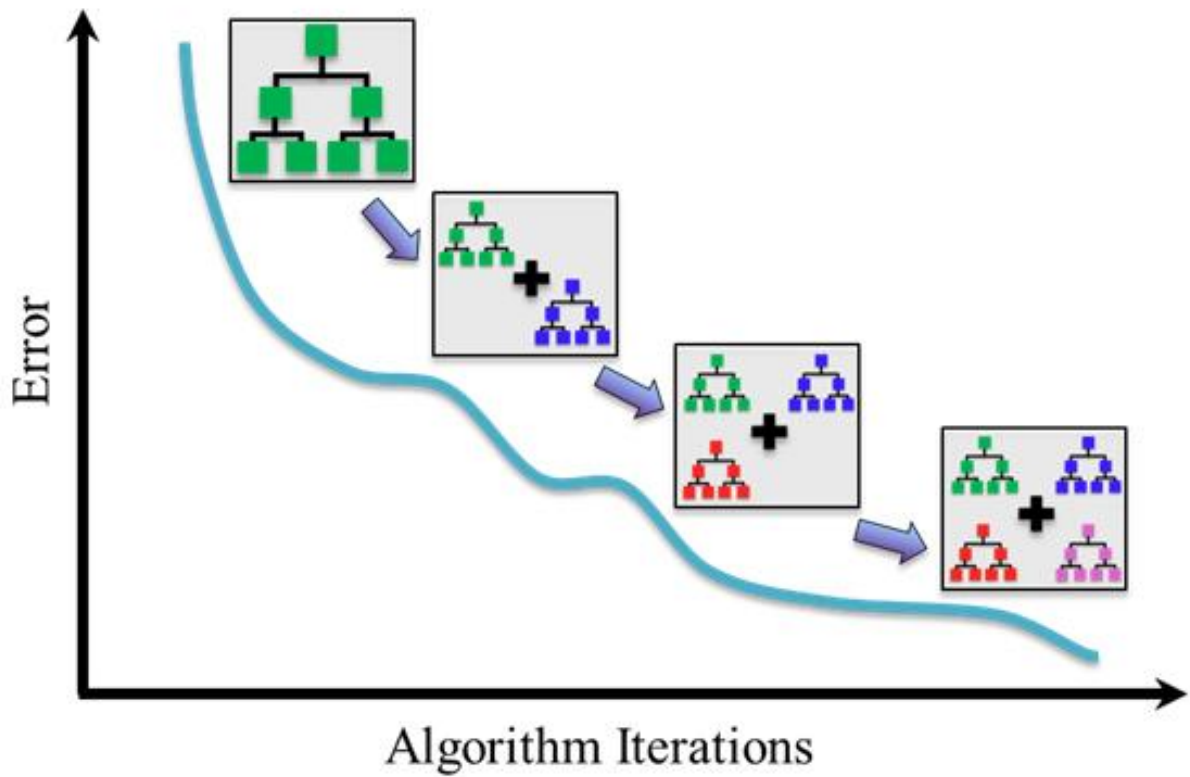

**Supplementary Figure A.9**
**Gradient Boosting.** A gradient boosting algorithm adds sub-models to the ensemble incrementally with the goal of minimizing the loss function (i.e. model error).

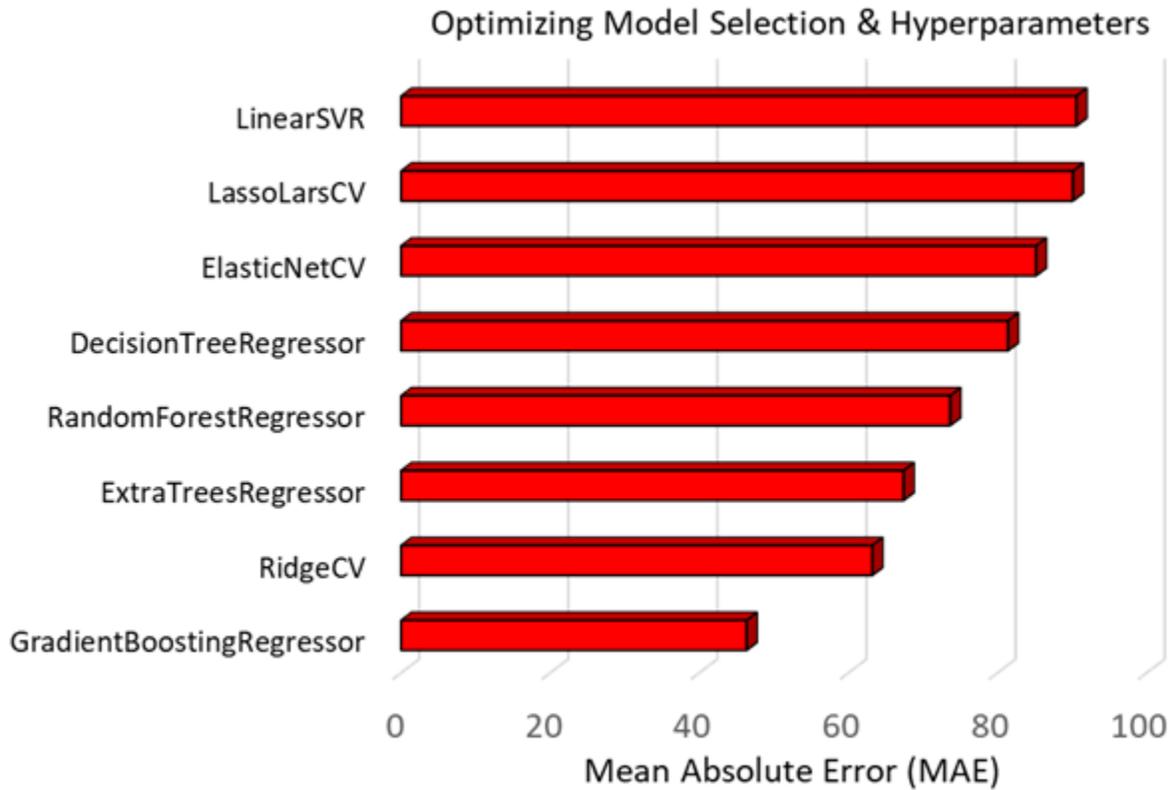

**Supplementary Figure A.10**
**Benchmarking models and hyperparameters.** The Automatminer software package uses a built-in genetic algorithm to test and optimize eight potential regression models and their hyperparameters. Each combination is scored using mean absolute error and the best performing model of each type is plotted above. The GradientBoostingRegressor model has the lowest mean absolute error and is thus selected.

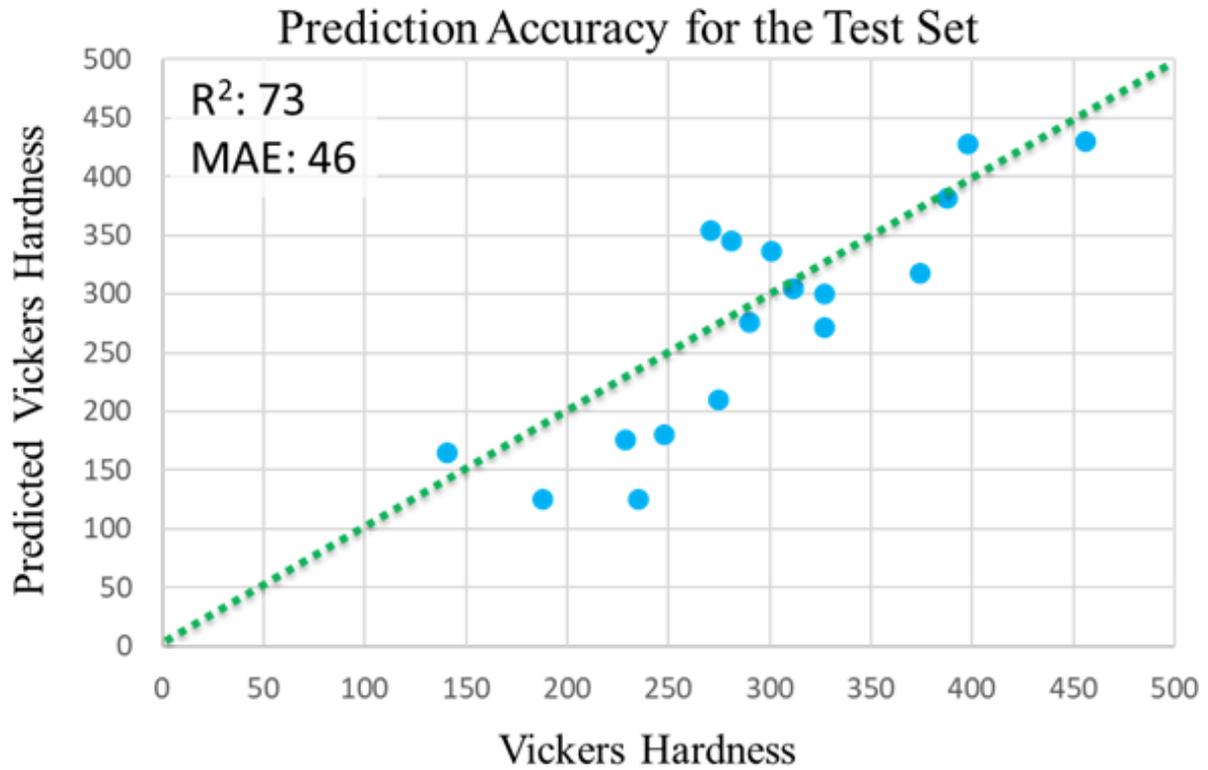

**Supplementary Figure A.11**
**ML model performance on the test data.** Predictions on holdout data demonstrate the model's effectiveness for exploring new composition space. Mean absolute error (MAE) and the coefficient of determination ($R^2$) are provided.

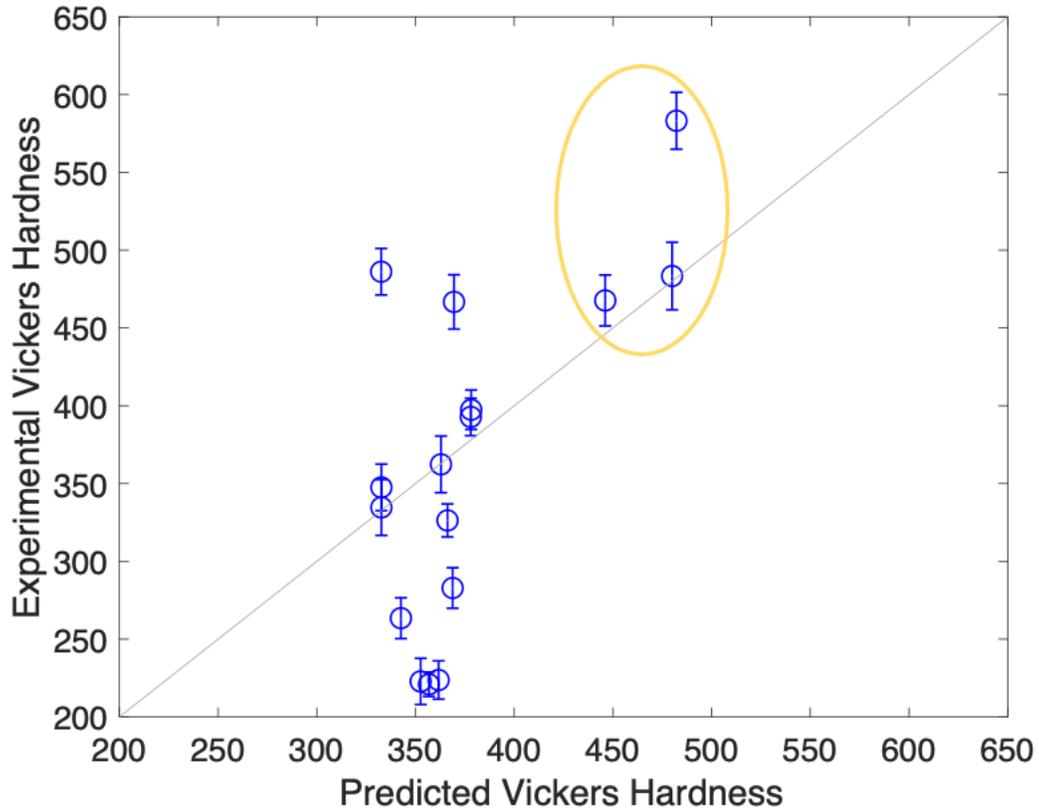

**Supplementary Figure A.12**
**ML model performance on the experimental nickel alloys.** Applying the model to the Ni-based alloys in the first sample ring, the ML model predicts 3 of the compositions, circled in yellow, will have higher hardness values than the other 13. Indeed, these are 3 of the hardest samples tested in this iteration and are among the 5 materials that have a higher Vickers hardness than the commercial materials in the dataset. This new data can be incorporated into the next training of the ML model, to improve subsequent predictions.

**Supplementary Table A.1**

**CALPHAD-predicted phases and phases detected in XRD for 16 In625-based compositions.**
CALPHAD predictions used the TCNI10 database.

| Sample # | Composition | Phases predicted (CALPHAD) 1100°C | Phases present (XRD) |
|---|---|---|---|
| 1 | In625 | Ni | Ni |
| 2 | In625 +3% Nb | Ni | Ni |
| 3 | In625 +6% Nb | Ni, 6% $Ni_3Ta$ | Ni, $Ni_3Ta$ |
| 4 | In625 +8% Nb | Ni, 16% $Ni_3Ta$, 2% Sigma | Ni, $Ni_3Ta$, Sigma |
| 5 | In625 +11% Nb | Ni, 26% $Ni_3Ta$, 6% Sigma | Ni, $Ni_3Ta$, Sigma |
| 6 | In625 +13% Nb | Ni, 36% $Ni_3Ta$, 13% Sigma | Ni, $Ni_3Ta$, Sigma |
| 7 | In625 +5% Mo | Ni | Ni |
| 8 | In625 +9% Mo | Ni, 5% Sigma | Ni |
| 9 | In625 +12% Mo | Ni, 13% Sigma | Ni, P phase |
| 10 | In625 +16% Mo | Ni, 17% Sigma, 5% P phase | Ni, Sigma, P phase |
| 11 | In625 +19% Mo | Ni, 17% Sigma, 13% P phase | Ni, Sigma, P Phase |
| 12 | In625 +3% Cr | Ni | Ni |
| 13 | In625 +6% Cr | Ni | Ni |
| 14 | In625 +10% Cr | Ni, 0.3% Sigma | Ni |
| 15 | In625 +13% Cr | Ni, 3% Sigma | Ni, Sigma |
| 16 | In625 +15% Cr | Ni, 7% Sigma | Ni, Sigma |

**Supplementary Table A.2**

**Distribution of Vickers hardness for the training and test data.** The maximum, minimum, mean, and standard deviation for each data set used to fit and evaluate the machine learning model. The test set is independent yet near-identically distributed to the training set.

|  | Training Data | Test Data |
| --- | --- | --- |
| **Minimum** | 125 | 125 |
| **Maximum** | 484 | 430 |
| **Mean** | 310 | 277 |
| **Standard Deviation** | 93.5 | 99.1 |

**Supplementary Table A.3**

**GradientBoostingRegressor hyperparameters.** Hyperparameter values selected for the GradientBoostingRegressor machine learning model. These are optimized via the genetic algorithm built into the Automatminer software package.

| Hyperparameter | Value |
| --- | --- |
| Number of Estimators | 500 |
| Alpha | 0.99 |
| Ccp Alpha | 0.0 |
| Criterion | Friedman_mse |
| Learning Rate | 0.1 |
| Loss | Huber |
| Max Depth | 5 |
| Max Features | 0.45 |
| Subsample | 0.9 |
| Tol | 0.0001 |
| Validation Fraction | 0.1 |

## Supplementary Table A.4

**Feature importance.** The top ten features for the ML model are listed. The avg and avg. deviation denote the composition-weighted average and average deviation, respectively, calculated over the vector of elemental values for each compound.

| Rank | Feature |
|------|---------|
| 1 | Avg Deviation Number of Valence Electrons |
| 2 | Solid Solution Strengthening Coefficient at 600K |
| 3 | Avg Deviation DFT volume per Atom at T=0K Ground State |
| 4 | Avg Deviation Number of d-Valence Electrons |
| 5 | Yang Omega – Mixing Thermochemistry |
| 6 | Phase Fraction of FCC_L12 Phase at 800K |
| 7 | Solidus Temperature |
| 8 | Phase Fraction of FCC_L12 Phase at 600K |
| 9 | Avg Deviation Melting Temperature |
| 10 | Avg Deviation Electronegativity |

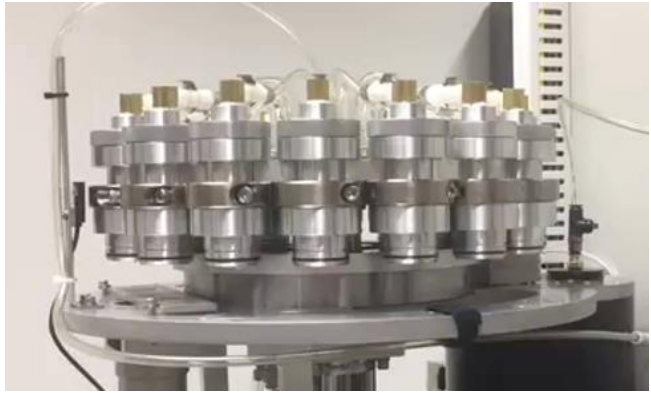

**Supplementary Movie A.1**

**Automated rotation of the Alloy Development Feeder (ADF, Formalloy).** Each of the 16 vials can contain a different alloy composition. The automated rotation of the ADF is controlled through the additive build program.

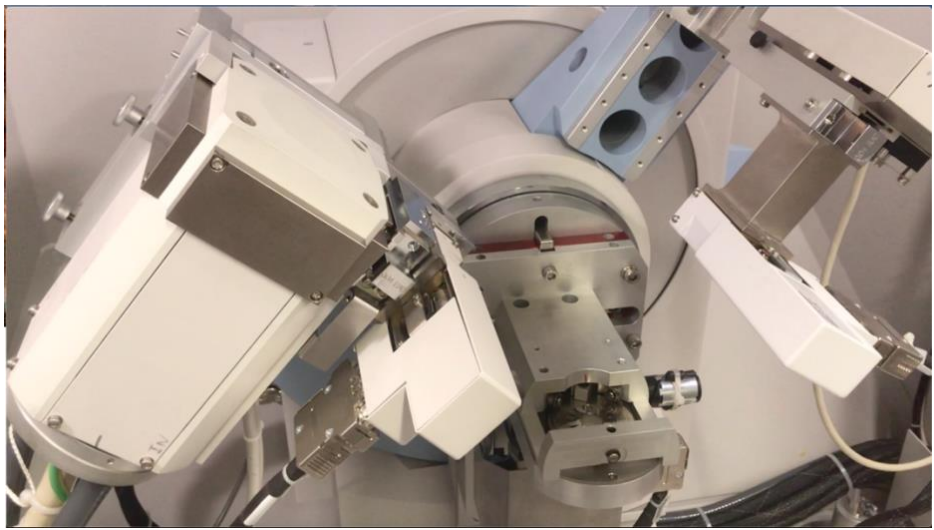

**Supplementary Movie A.2**

**Automated rotation of the sample about its center axis in the Panalytical X'Pert Pro XRD.** Green circle indicates sample location. Sample rotates every 10 seconds to illustrate automated rotation ability. Data collection time per sample is approximately 10 minutes.

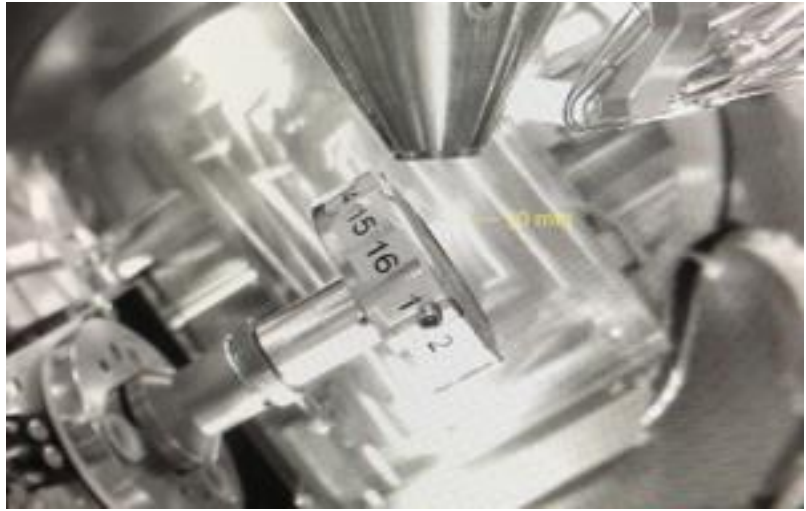

**Supplementary Movie A.3**
**Automated rotation of the sample about its center axis in the SEM, side view.** Sample rotation enables beam geometry to stay constant when the sample is rotated during automated data collection.

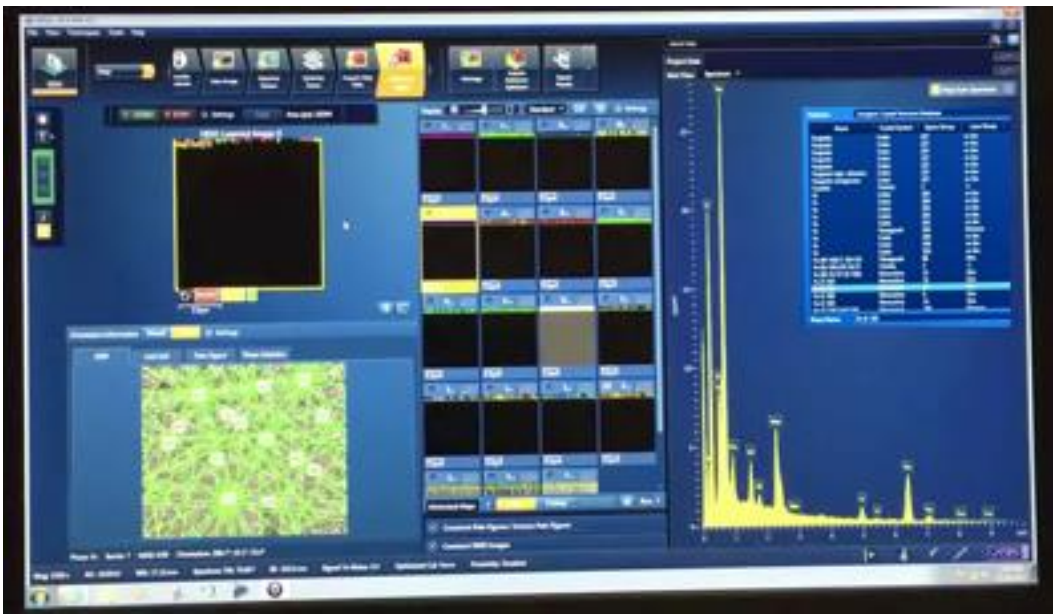

**Supplementary Movie A.4**
**Automated SEM-EDS/EBSD data collection.** Data collection time per sample is approximately 1 minute.

# Appendix B: CALPHAD Data

High Throughput Rapid Experimental Alloy Development (HT-READ)

Olivia F. Dippo, Kevin R. Kaufmann, Kenneth S. Vecchio

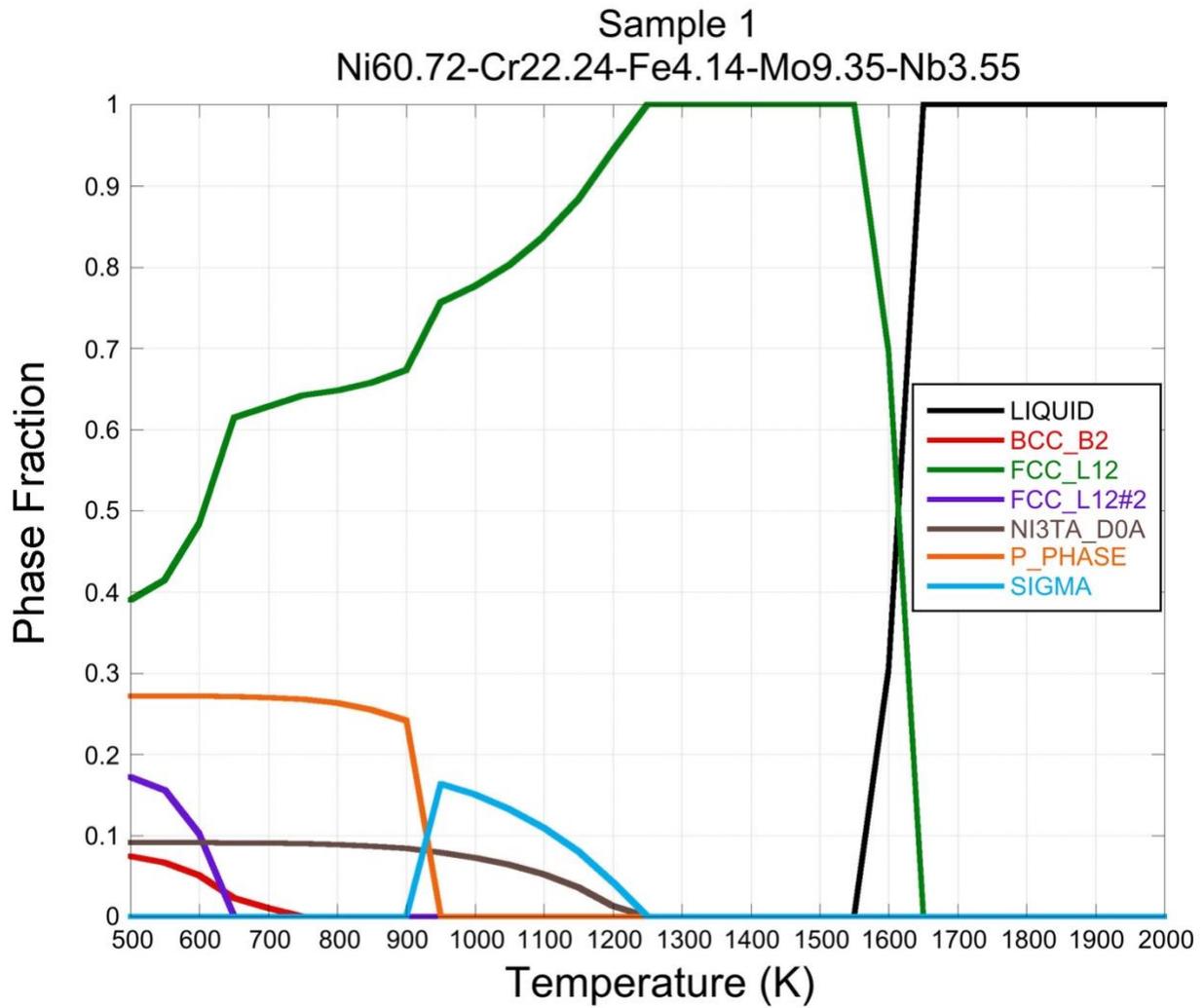

**Supplementary Figure B.1**
**Phase evolution diagram for sample 1.** CALPHAD data for sample 1: In625

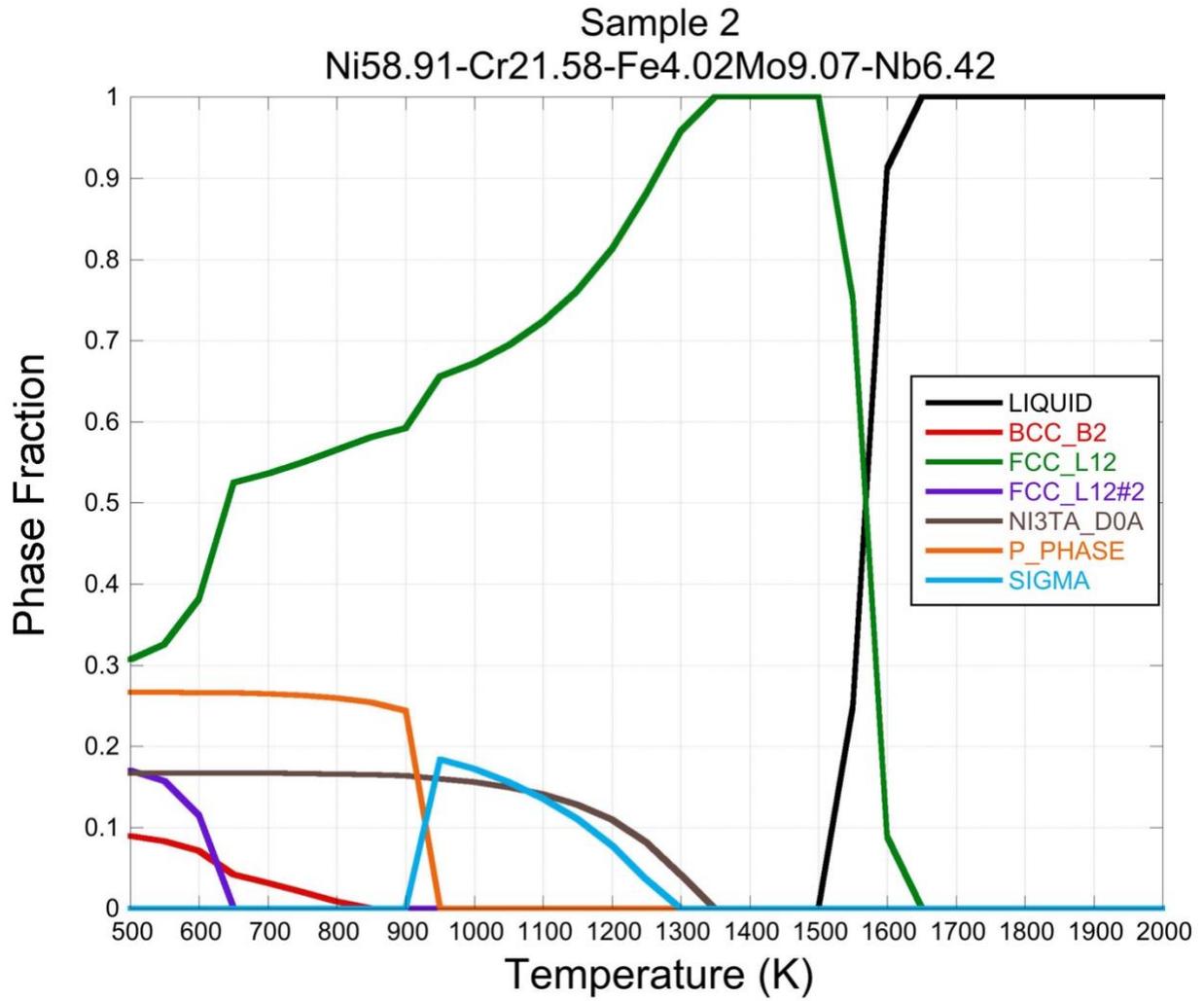

**Supplementary Figure B.2**
**Phase evolution diagram for sample 2.** CALPHAD data for Sample 2: In625 + 3% Nb

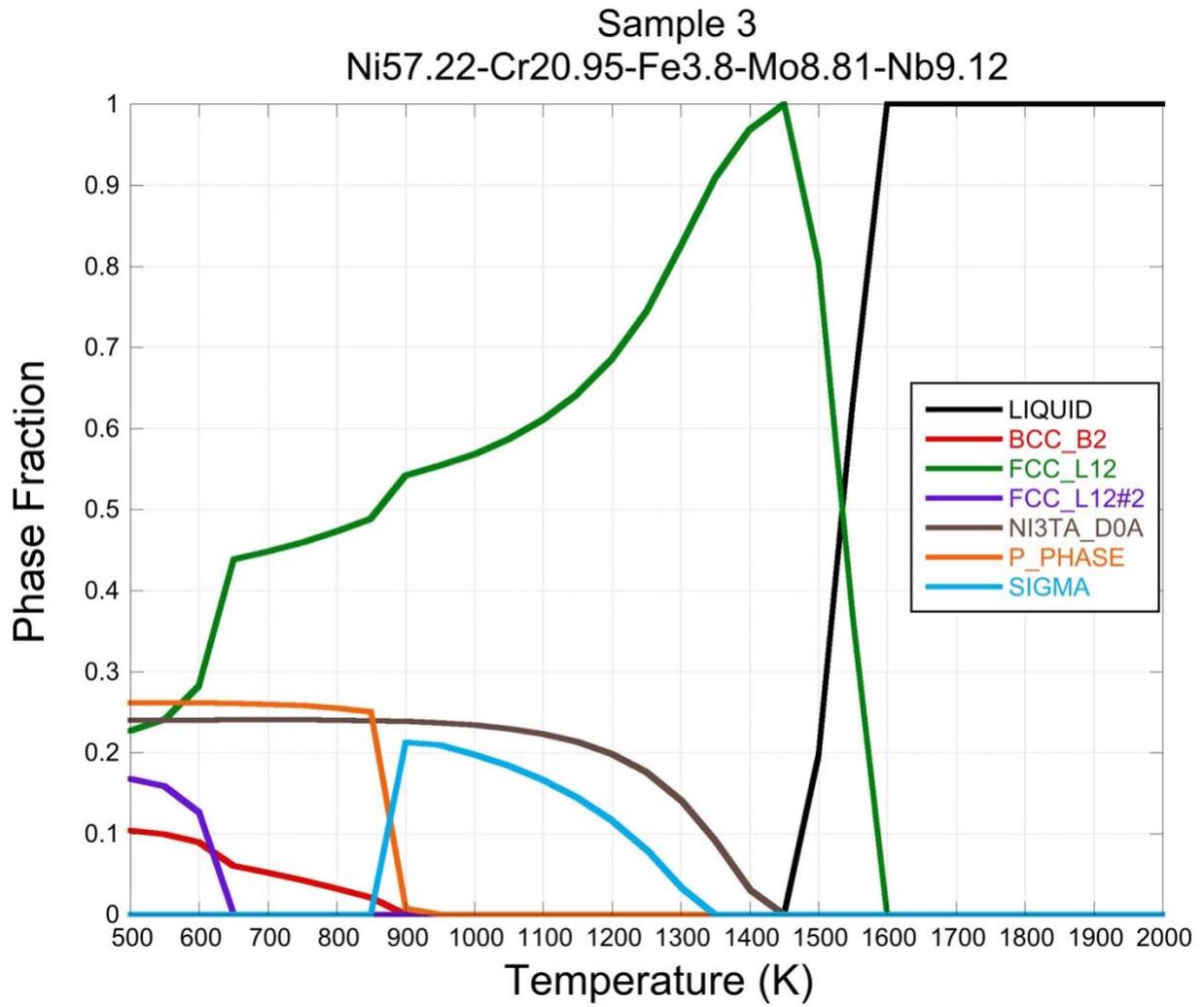

**Supplementary Figure B.3**
**Phase evolution diagram for sample 3.** CALPHAD data for Sample 3: In625 + 6% Nb

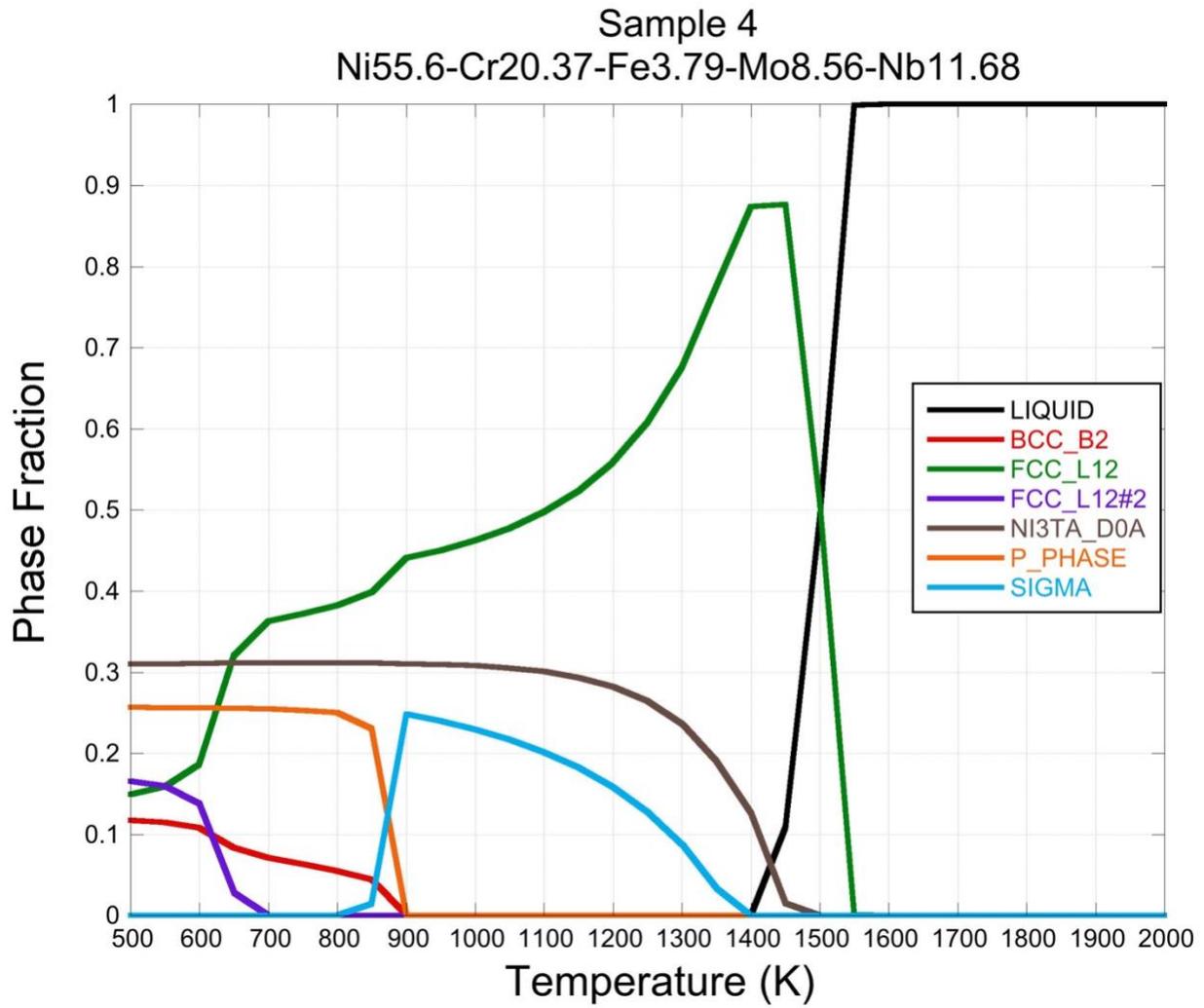

**Supplementary Figure B.4**
**Phase evolution diagram for sample 4.** CALPHAD data for Sample 4: In625 + 8% Nb

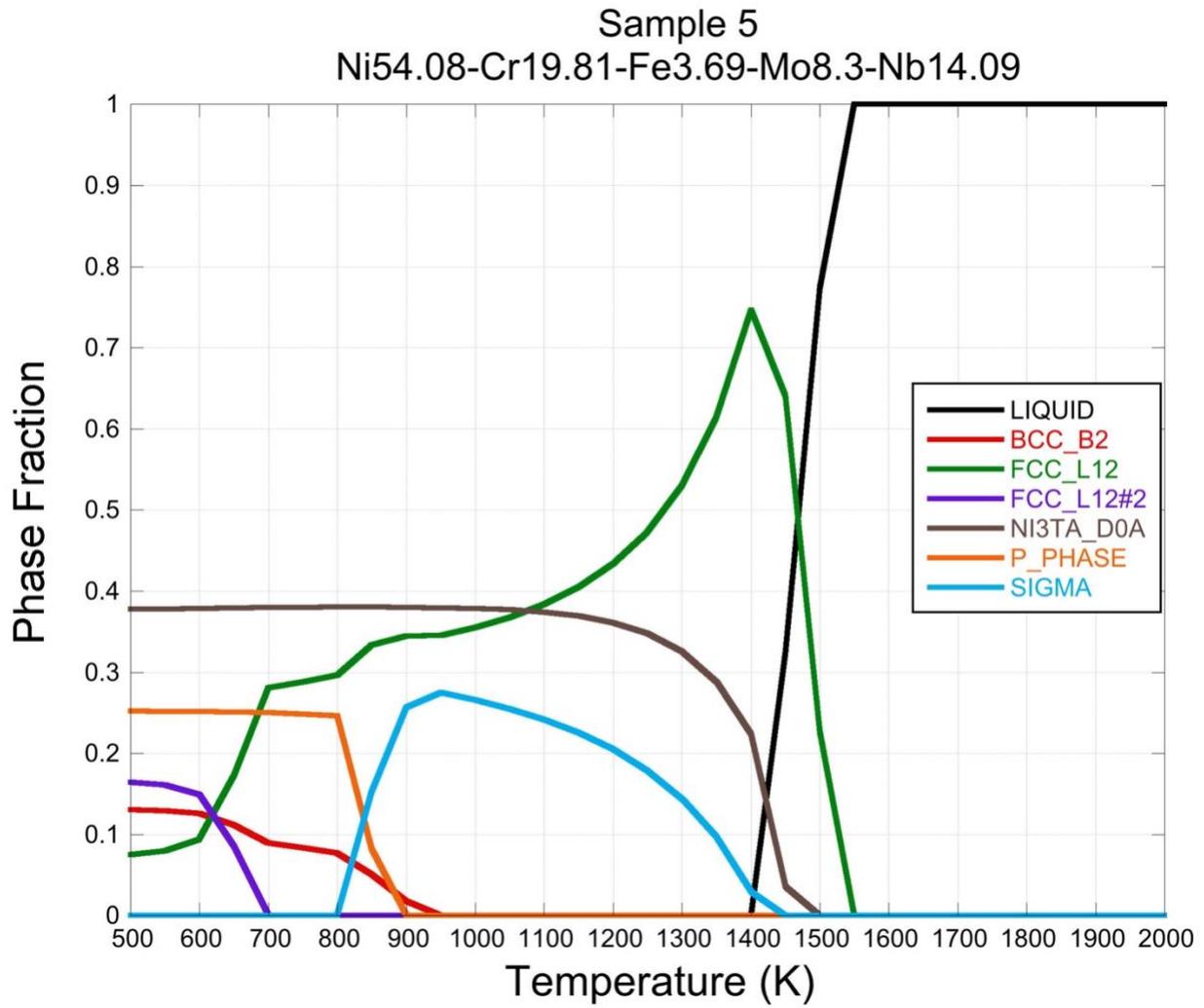

**Supplementary Figure B.5**
**Phase evolution diagram for sample 5.** CALPHAD data for Sample 5: In625 + 11% Nb

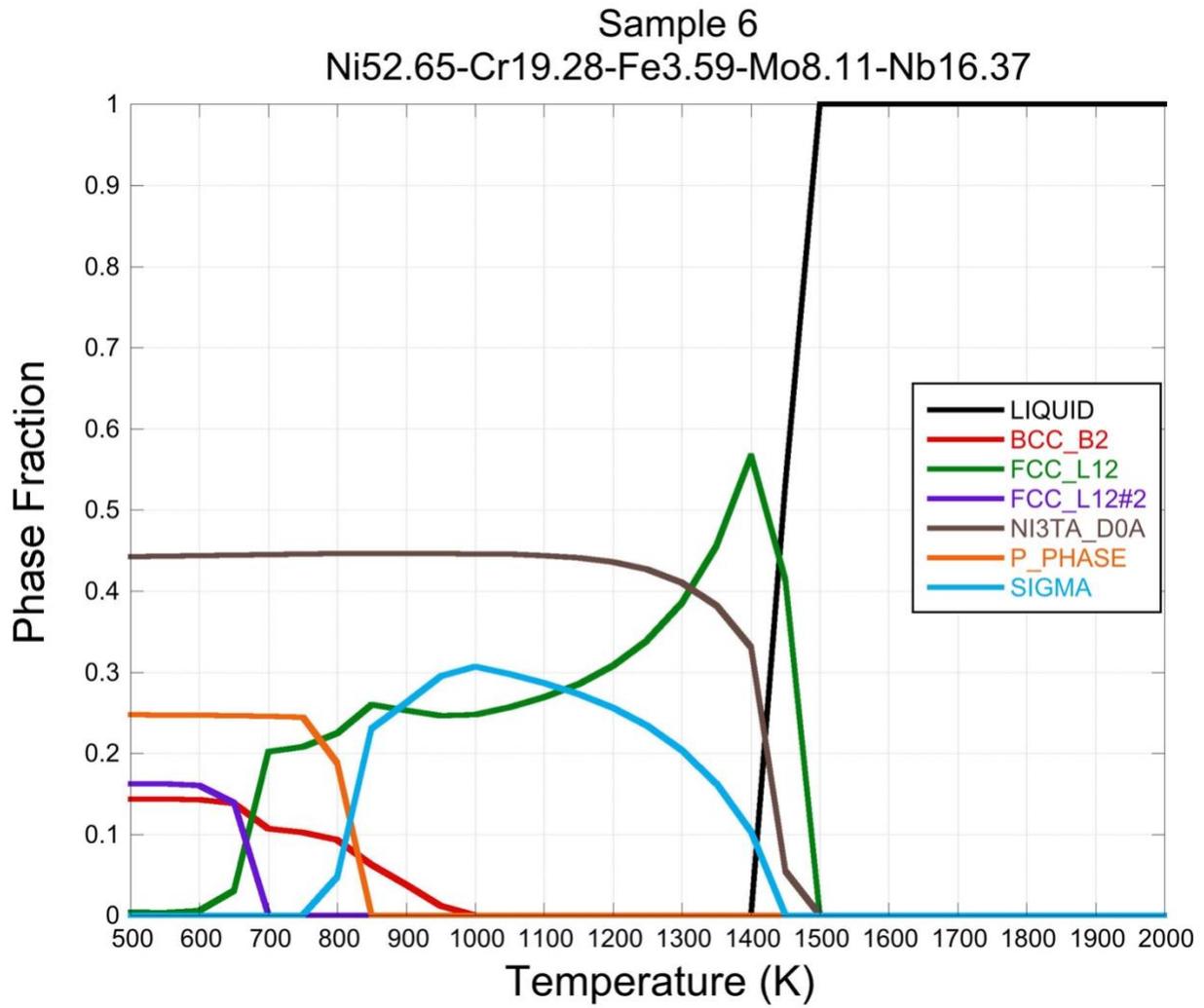

**Supplementary Figure B.6**
**Phase evolution diagram for sample 6.** CALPHAD data for Sample 6: In625 + 13% Nb

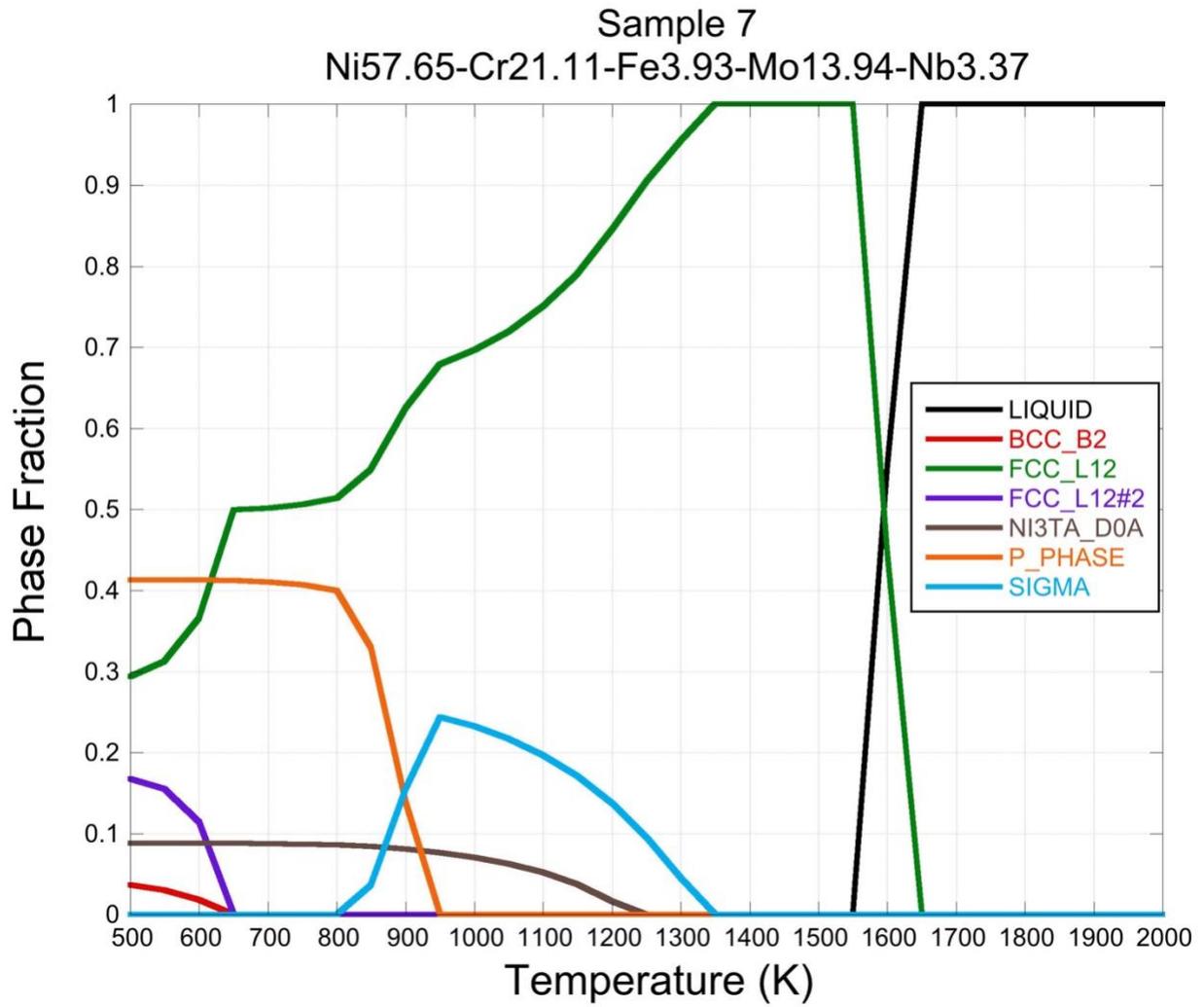

**Supplementary Figure B.7**
**Phase evolution diagram for sample 7** CALPHAD data for Sample 7: In625 + 5% Mo

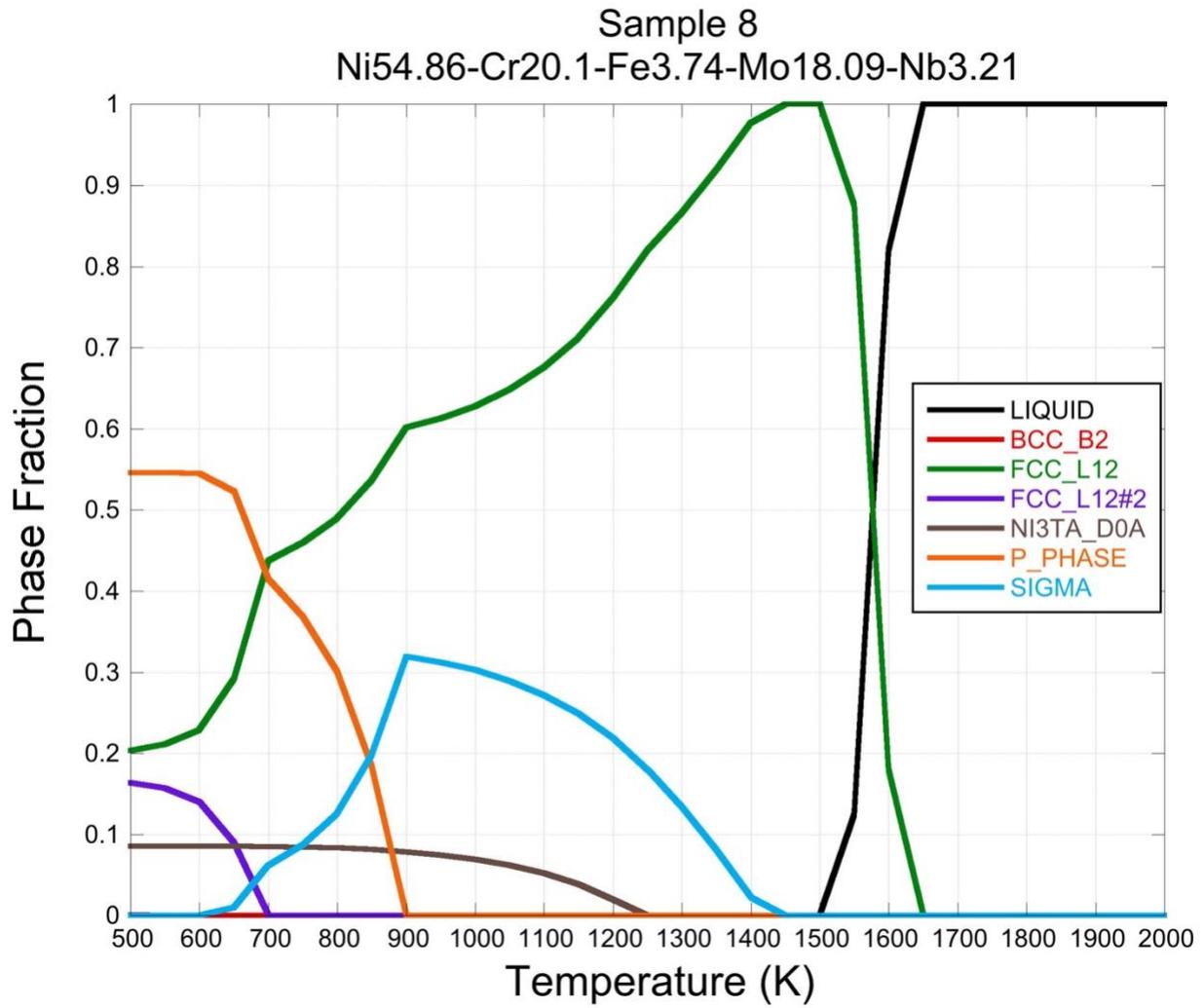

**Supplementary Figure B.8**
**Phase evolution diagram for sample 8.** CALPHAD data for Sample 8: In625 + 9% Mo

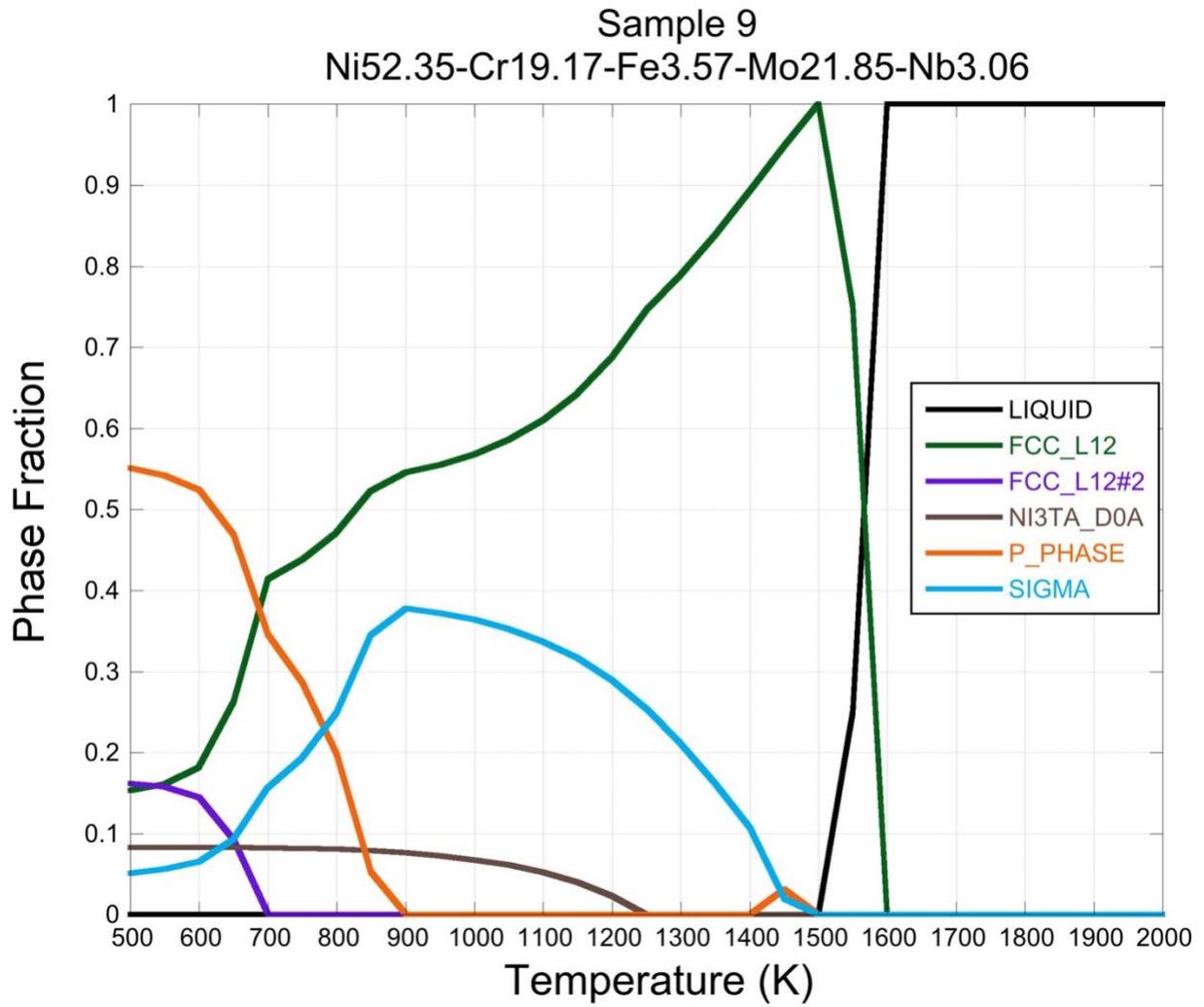

**Supplementary Figure B.9**
**Phase evolution diagram for sample 9.** CALPHAD data for Sample 9: In625 + 12% Mo

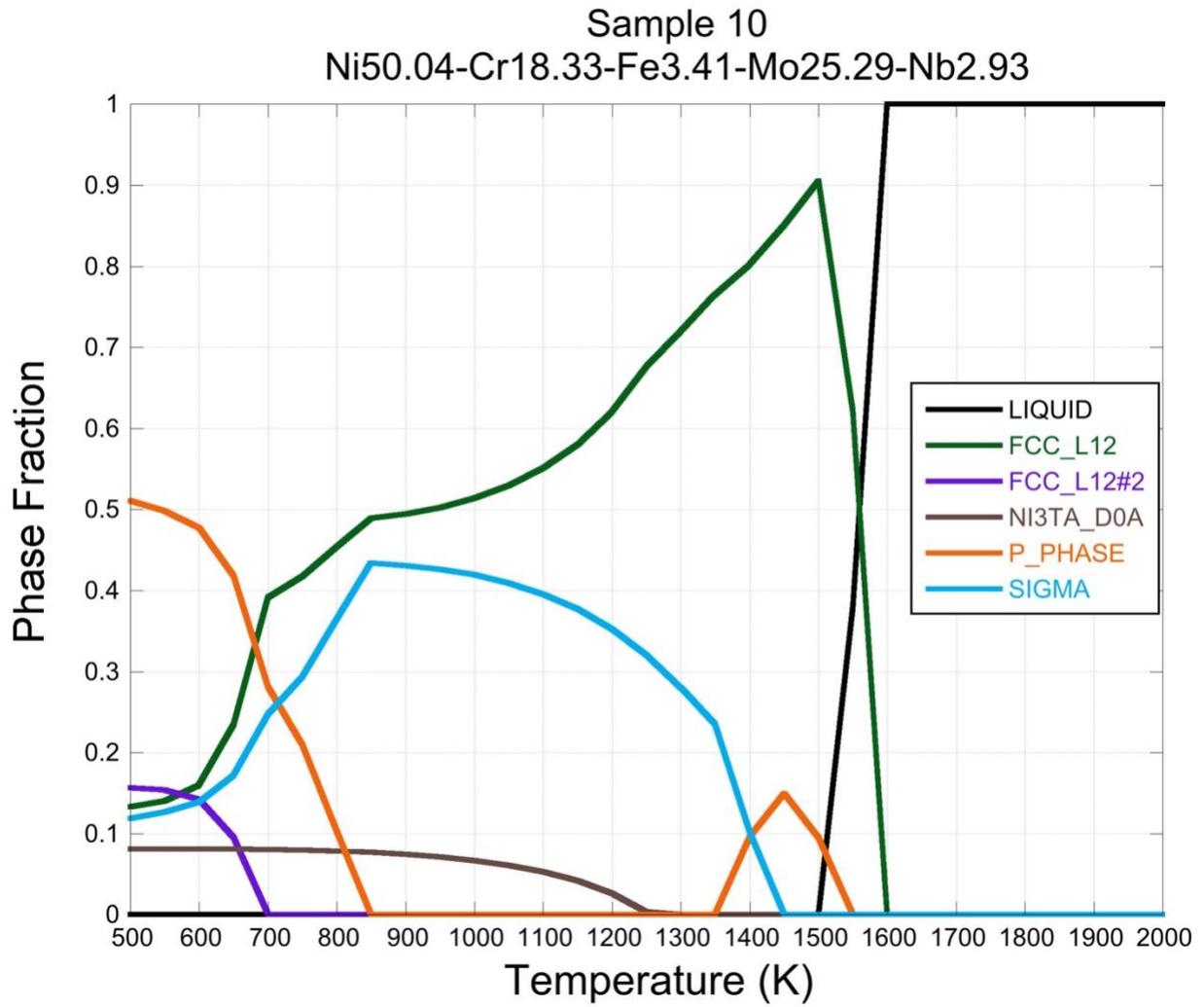

**Supplementary Figure B.10**
**Phase evolution diagram for sample 10.** CALPHAD data for Sample 10: In625 + 16% Mo

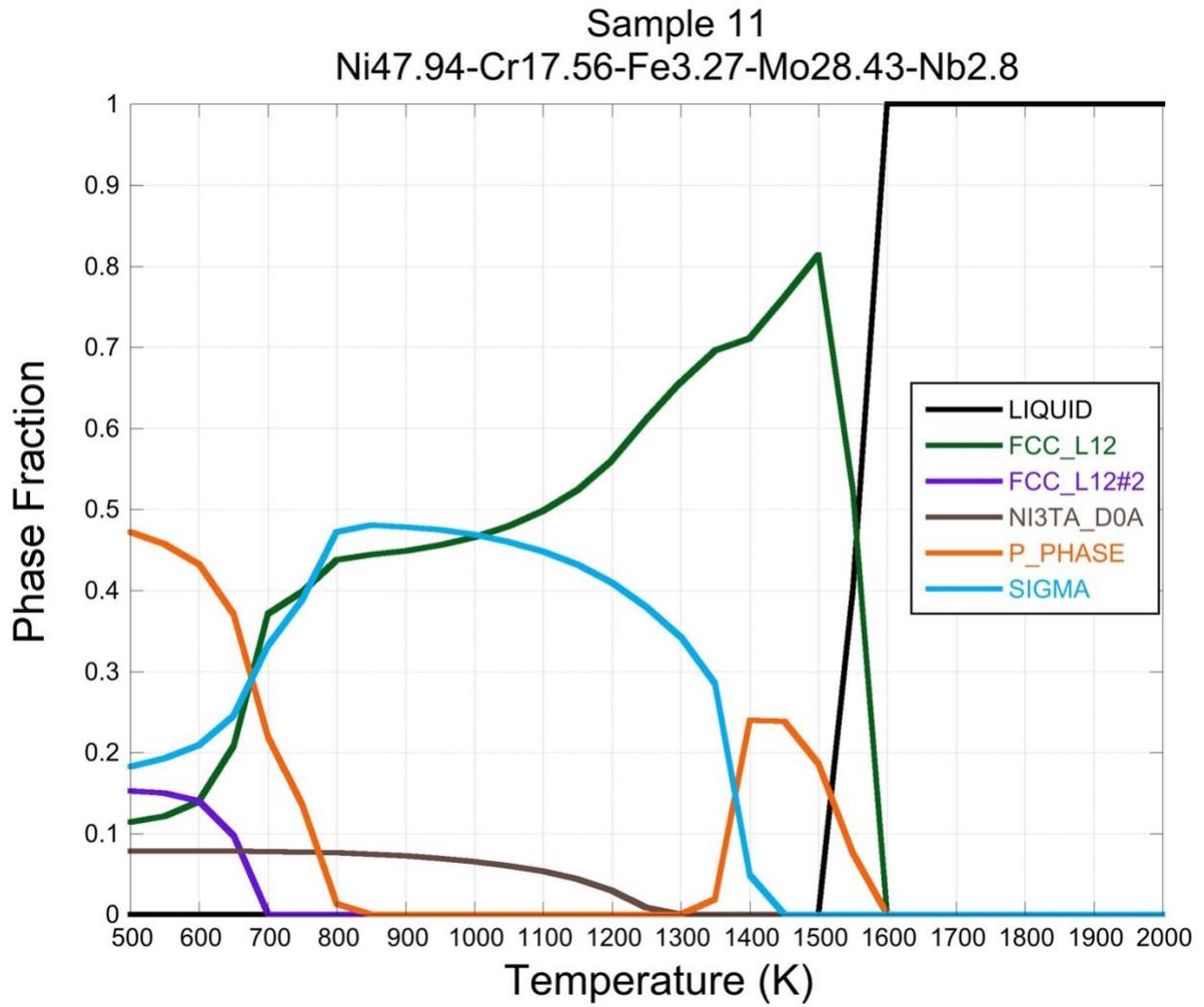

**Supplementary Figure B.11**
**Phase evolution diagram for sample 11** CALPHAD data for Sample 11: In625 + 19% Mo

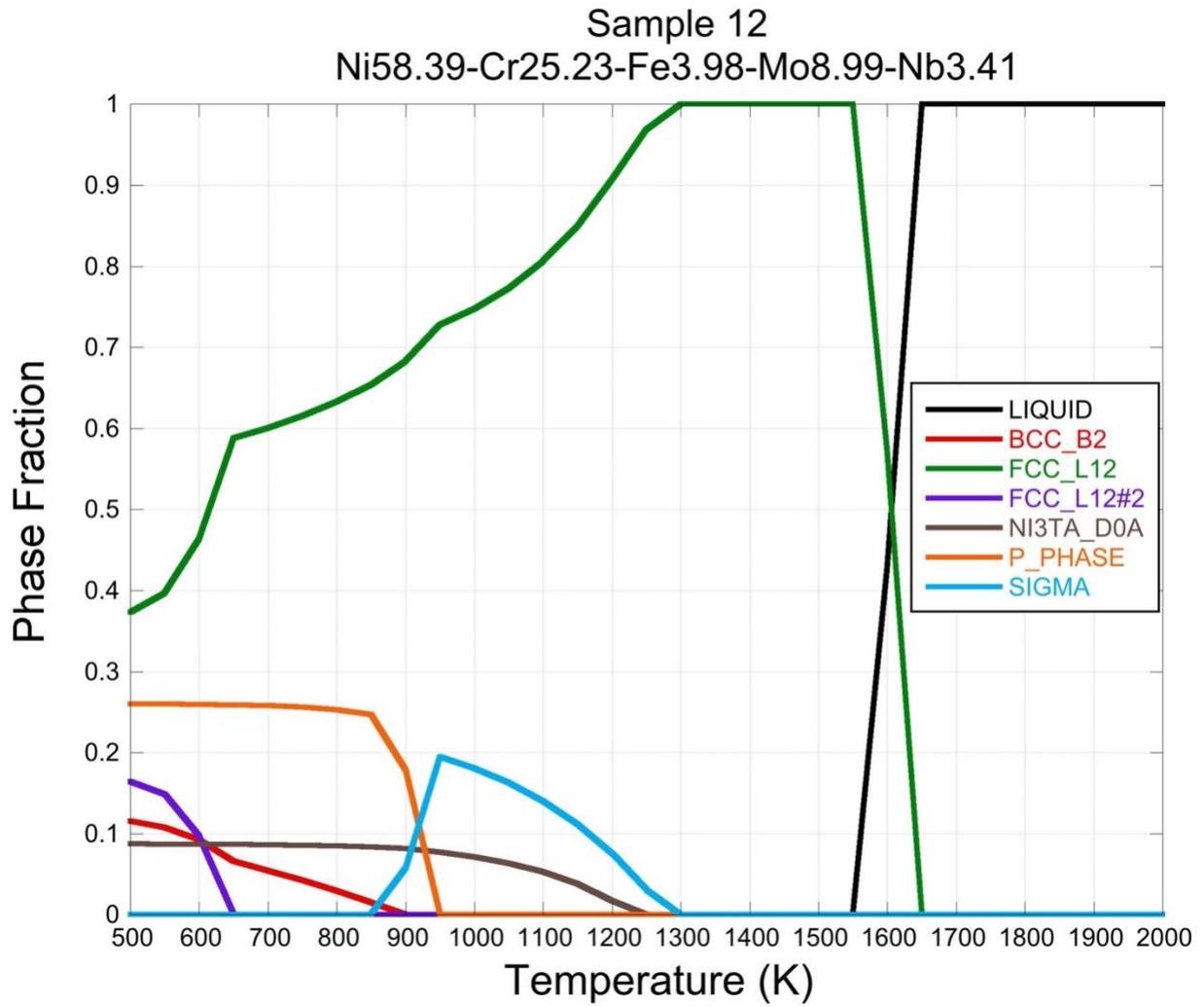

**Supplementary Figure B.12**
**Phase evolution diagram for sample 12.** CALPHAD data for Sample 12: In625 + 3% Cr

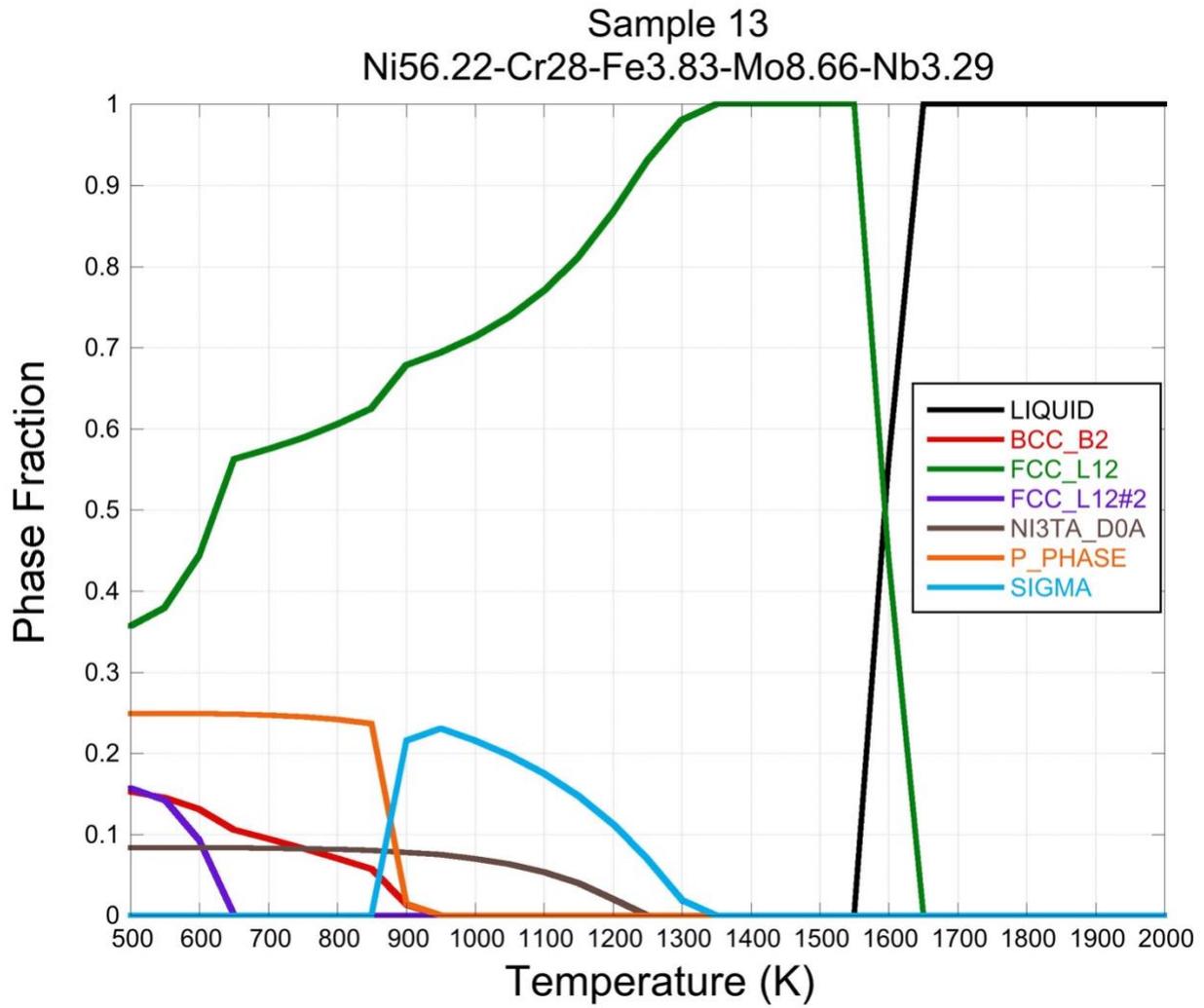

**Supplementary Figure B.13**
**Phase evolution diagram for sample 13.** CALPHAD data for Sample 13: In625 + 6% Cr

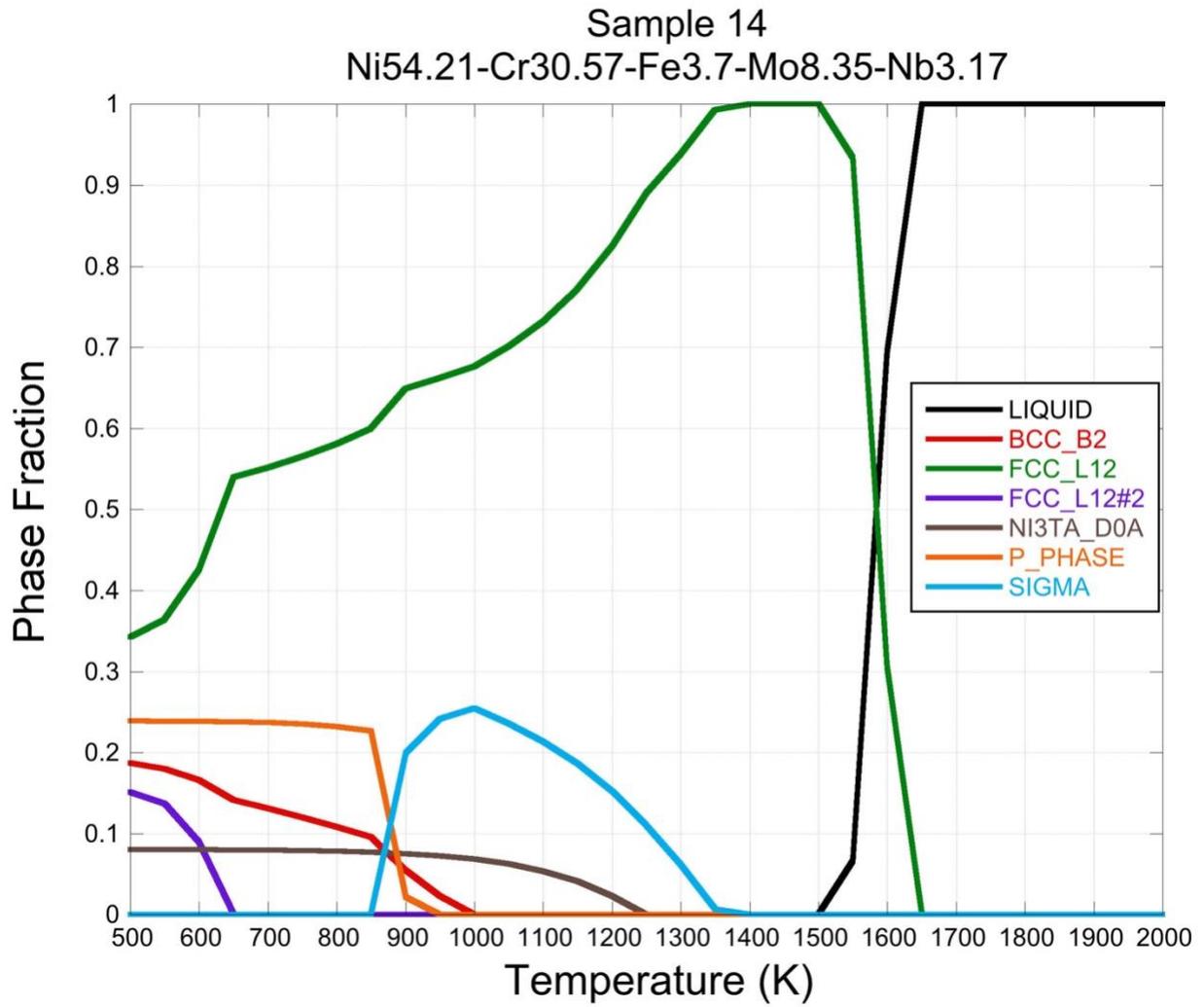

**Supplementary Figure B.14**
**Phase evolution diagram for sample 14.** CALPHAD data for Sample 14: In625 + 10% Cr

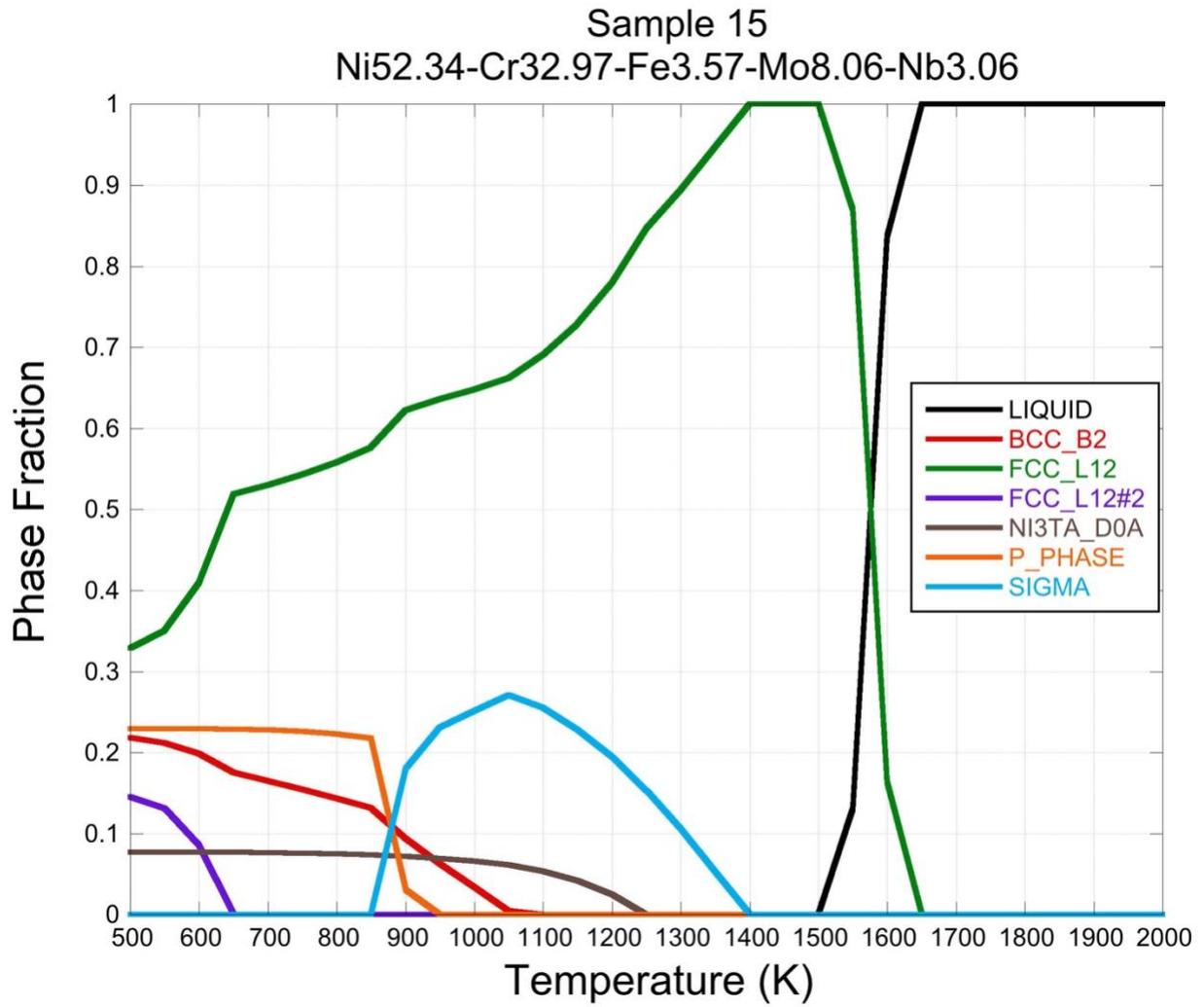

**Supplementary Figure B.15**
**Phase evolution diagram for sample 15.** CALPHAD data for Sample 15: In625 + 13% Cr

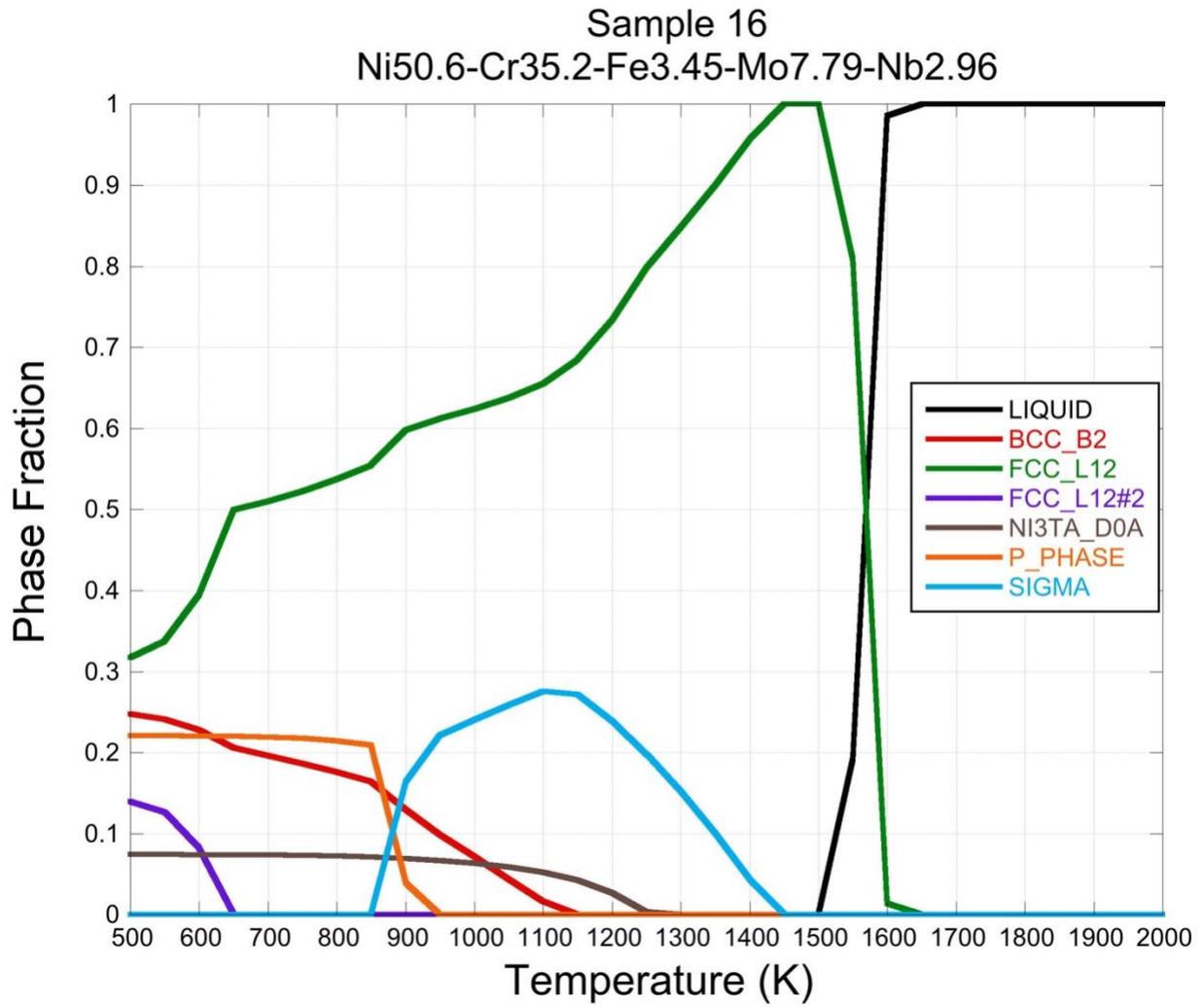

**Supplementary Figure B.16**
**Phase evolution diagram for sample 16.** CALPHAD data for Sample 16: In625 + 15% Cr